\documentclass{article}
\usepackage{graphicx} 
\usepackage{hyperref}
\usepackage{tikz}
\usepackage[utf8]{inputenc}
\usepackage[english]{babel}
\usepackage{amsmath,amsfonts,amssymb,amsfonts, epsfig, float, enumerate, color, listings, fancyhdr}
\usepackage[margin=2.5cm]{geometry}
\usepackage{graphicx}

\usepackage{float}
\usepackage{array}
\usepackage{tocloft}
\usepackage{booktabs}
\usepackage{titlesec}
\usepackage{listings} 
\usepackage{pdfpages}
\usepackage{cite}
\usepackage{tabularx}
\usepackage{hyperref}
\usepackage{caption}
\usepackage{comment}
\usepackage{subfig}
\usepackage{appendix} 
\usepackage{tabularray}
\usepackage{multirow}
\usepackage{authblk}
\usepackage[normalem]{ulem}

\title{Investigation of Low-Energy Particle Remnants in High-Energy Collisions at the LHC with a Skipper-CCD detector}

\author[1]{Brenda A. Cervantes-Vergara \thanks{bcervant@fnal.gov}}
\author[1,2,3]{Santiago E. Perez\thanks{santiep.137@gmail.com}}
\author[1,4]{Nicola Bacchetta}
\author[5]{Nuria Castelló-Mor}
\author[1]{Juan Estrada}
\author[5]{Marcos Fernandez Garcia}
\author[1]{Petra Merkel}
\author[5]{María Pérez Martínez}
\author[2,3]{Dario Rodrigues}
\author[1]{Javier Tiffenberg}
\author[5]{Rocío Vilar Cortabitarte}

\affil[1]{Fermi National Accelerator Laboratory, 
          Batavia, IL, United States}
\affil[2]{Departamento de Física, FCEN, Universidad de Buenos Aires, Buenos Aires, Argentina} 
\affil[3]{Instituto de Física de Buenos Aires (IFIBA), CONICET and Universidad de Buenos Aires, Buenos Aires, Argentina}
\affil[4]{Istituto Nazionale di Fisica Nucleare (INFN), Padova, Italy }
\affil[5]{Instituto de Fisica de Cantabria, Santander, Spain }
\begin{document}

\maketitle

\begin{abstract}
We deployed MOSKITA $\sim$33~m away from the CMS collision point, the first skipper-CCD detector probing low-energy particles produced in high-energy collisions at the Large Hadron Collider (LHC). In this work, we search for beam-related events using data collected in 2024 during beam-on and beam-off periods. The dataset corresponds to integrated luminosities of 113.3~fb$^{-1}$ and 1.54~nb$^{-1}$ for the proton-proton and Pb–Pb collision periods, respectively. We report observed event rates in a model-independent framework across two ionization regions: $\leq20e^-$ and $>20e^-$. For the low-energy region, we perform a likelihood analysis to test the null hypothesis of no beam-correlated signal. We found no significant correlation during proton-proton and Pb-Pb collisions. For the high-energy region, we present the energy spectra for both collision periods and compare event rates for images with and without luminosity. We observe a slight increase in the event rate following the Pb-Pb collisions, coinciding with a rise in the single-electron rate, which will be investigated in future work. Using the low-energy proton-proton results, we place 95\% C.L. constraints on the mass–millicharge parameter space of millicharged particles. Overall, the results in this work demonstrate the viability of skipper-CCD technology to explore new physics at high-energy colliders and motivate future searches with more massive detectors.
\end{abstract}

\tableofcontents

\section{Introduction}
The search for new physics at the Large Hadron Collider (LHC) has traditionally focused on high-energy signatures predicted by various extensions of the Standard Model~\cite{Kopp2013, Schramm:2022ngt, Biswas:2022tcw}. However, a complementary and increasingly promising approach involves probing low-energy particle remnants produced in LHC's high-energy collisions. This unexplored regime may reveal subtle signals from weakly interacting or hidden-sector particles, such as millicharged particles (mCPs), which are predicted in well-motivated theoretical scenarios involving kinetic mixing between Standard Model and hidden photons~\cite{MCPdarkPhoton}.

The skipper-CCD technology has already demonstrated world-leading sensitivity in the search for such particles. Its unique ability to resolve individual ionization electrons, combined with its low-energy threshold~\cite{Tiffenberg_2017}, enables the detection of faint signatures like those expected from mCP interactions. The SENSEI experiment at the MINOS cavern at Fermilab~\cite{SENSEI2020} used this technology to set world-leading constraints on mCPs in the sub-GeV mass range using the NuMI beam~\cite{mCP_by_SENSEI}. Also, the CONNIE and Atucha-II collaborations, with skipper-CCD detectors operating at nuclear power plants in South America, have placed the strongest exclusion limits for mCPs in the 1~eV to 700~MeV mass range~\cite{mCP_from_reactors}. However, the production of heavier mCPs is highly suppressed in these environments.

The 13.6~TeV proton-proton collisions at the LHC open a new window to probe GeV-mass mCPs. The milliQan experiment has recently placed world-leading limits for mCPs with masses above 0.45~GeV~\cite{milliqan2025}. Similar efforts are ongoing at several locations along LHC~\cite{MoEDAL2024, FASER2025, FORMOSA2025, Flare2023}, including both upgrades to existing detectors and new proposals for the High-Luminosity LHC (HL-LHC).

To contribute to this program, we deployed MOSKITA (MObile SKIpper Testing Apparatus) at the LHC, the first skipper-CCD detector sensitive to low-energy ionization signals from high-energy collisions. MOSKITA enables the characterization of both environmental and beam-related ionization backgrounds, as well as the search for anomalous excesses that could indicate new physics.

In this work, we present the first results from MOSKITA, using data collected in 2024 during proton-proton collisions, Pb-Pb collisions, and beam-off periods. We provide details on the detector installation, performance, data acquisition, and analysis pipeline. We report the observed event rates in a model-independent framework across two ionization regions: $\leq20e^-$ and $>20e^-$. For the low-energy region, we report p-values and 95\% confidence level (C.L.) upper limits on the null hypothesis of no beam-correlated events, based on a profile likelihood analysis. We found no significant correlation during proton-proton and Pb-Pb collisions. For the high-energy region, we present the energy spectra for both collision periods, comparing rates for images with and without luminosity. We find an increase in the high-energy event rate following the Pb-Pb collisions, coinciding with a rise in the single-electron rate. Finally, using the low-energy results for the proton-proton collision period, we place constraints on the mass-millicharge parameter space for mCPs at a 95\% C.L., and project sensitivities for scenarios with a larger detector, reduced background, and higher luminosity.

\section{MOSKITA at the LHC}
\subsection{Experimental setup}
MOSKITA is a dedicated setup that has been previously used for testing skipper-CCDs for the Oscura experiment at FNAL~\cite{OscuraSensors2023, PM2023}. It consists of a stainless steel T-shaped vacuum vessel, placed inside a 2-inch-thick lead shield, which can accommodate a CCD sensor. All the hardware needed for its operation is located on a 1.2~m$^2$ aluminum pallet.

\begin{figure}[ht!]
\centering
    \includegraphics[height=8cm]{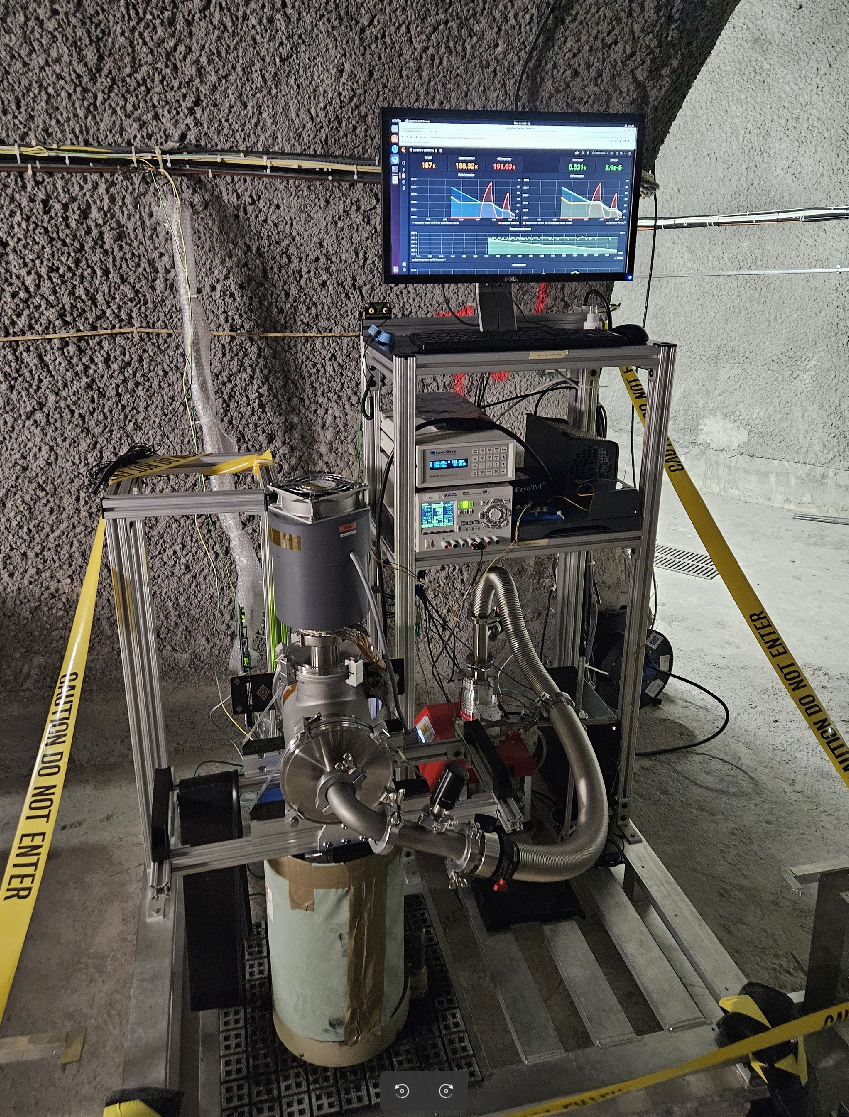}\hspace{1cm}
    \includegraphics[height=7.5cm]{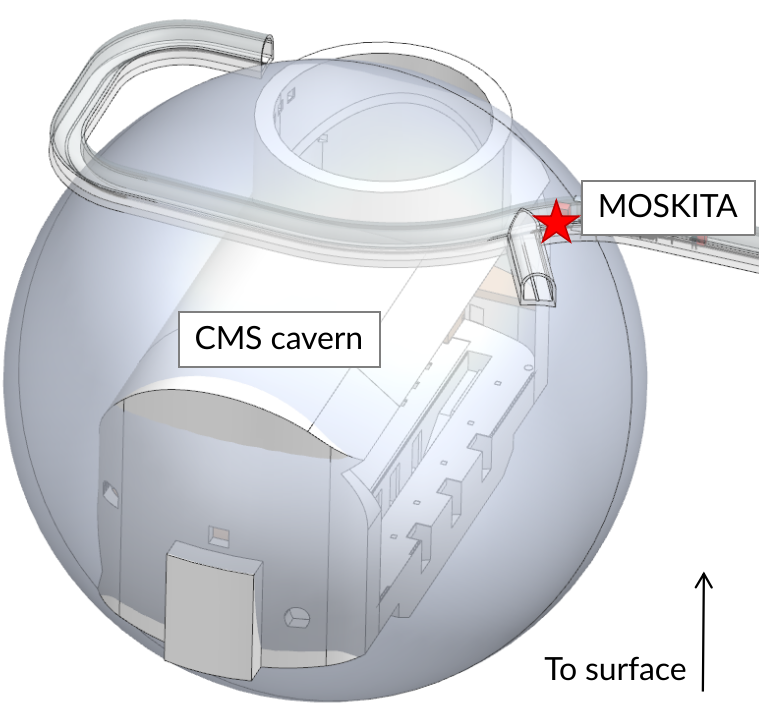}
    \caption{MOSKITA at the CMS drainage gallery (left); MOSKITA's location relative to CMS (right).}
    \label{fig:moskita-setup}
\end{figure}

MOSKITA was shipped to CERN and installed near LHC Interaction Point 5 in the CMS drainage gallery $\sim$70~m underground. This gallery, in which the milliQAN experiment is located~\cite{milliQAN2021, milliQAN2020}, is $\sim$33~m away from the interaction point, separated by $\sim$17~m of rock that shields the site from high-energy particles produced in collisions. According to the CMS coordinate system~\cite{CMS2008}, MOSKITA is placed at an azimuthal angle $\phi$ of $43^{\circ}$ and pseudorapidity $\eta$ of 0.1. Figure~\ref{fig:moskita-setup} shows a picture of MOSKITA installed at the CMS drainage gallery (left) and a diagram indicating the setup's position relative to the CMS detector (right).

We use as detector a 6.29~MPix p-channel skipper-CCD, designed by the Lawrence Berkeley National Laboratory (LBNL) and fabricated at Teledyne-DALSA for the SENSEI experiment~\cite{SENSEIsnolab2025}. The sensor's active area is an array of 1024 rows and 6144 columns of $15\times15~\mu$m$^2$ pixels and has a thickness of 675~$\mu$m, which gives a total mass of $\sim$2.2~grams. This sensor has four floating-gate amplifiers, one in each corner, that allow to perform multiple non-destructive measurements of the charge in each pixel during readout ($N_{\rm smp}$). The sensor's packaging was done at FNAL. In this process, the sensor and a copper-Kapton flexible cable are glued to a silicon substrate and wirebonded; then, the whole assembly is placed inside a copper tray.

Once MOSKITA was fully assembled at the LHC in March 2024, we installed the packaged sensor inside the vessel. During operations, the vessel holds a vacuum of $\sim$10$^{-4}$~ Torr, and the sensor is cooled down to $\sim$140K using a Sunpower CryoTel cryocooler to minimize thermally-generated charge. We use a Low-Threshold Acquisition board~\cite{LTA2021} to bias and read out the sensor. The temperature is monitored at two locations: on the sensor's copper tray (CCD temperature) and on the cryocooler's cold head, next to the heater used to keep the sensor's temperature constant (heater temperature). The pressure is monitored using a MicroPirani vacuum gauge.

\begin{figure}[ht!]
\centering
    \includegraphics[width=\linewidth, angle=180]{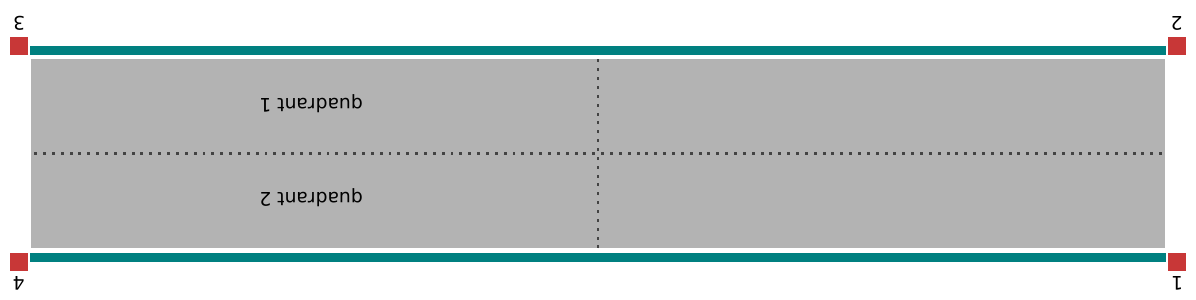}
    \caption{Diagram of the CCD installed in MOSKITA at the LHC. The two serial registers and the four output stages (not to scale) are highlighted in blue and red, respectively. The useful quadrants, when the CCD is read out through all four amplifiers, are labeled as quadrants 1 and 2 in the figure.}
    \label{fig:ccddiagram}
\end{figure}

\begin{figure}[ht!]
\centering
    \scalebox{-1}[1]{\includegraphics[width=0.95\linewidth]{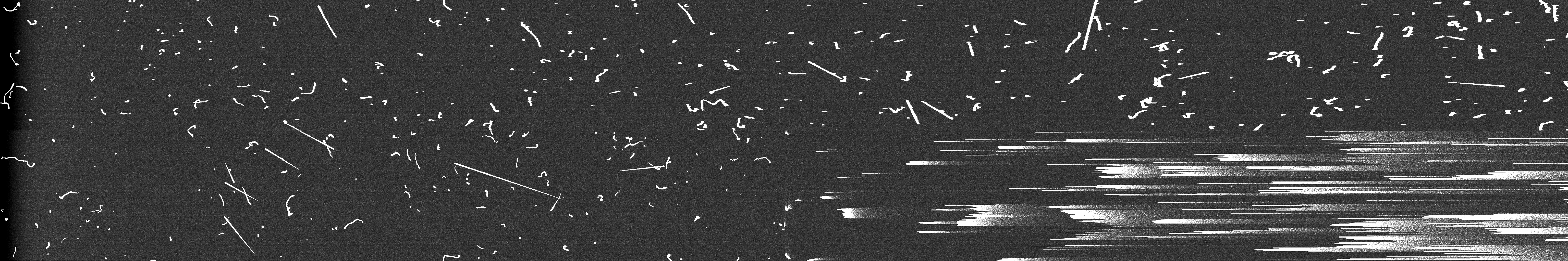}}
    \caption{Image taken in MOSKITA at the LHC, rearranged as in Figure~\ref{fig:ccddiagram}. It is a 3.5-hour exposure with a 15-minute readout through output stages 3 and 4. The charge transfer problem in the half of the serial register next to output stage 2 is clear.}
    \label{fig:imagemoskita}
\end{figure}

The sensor's performance was previously evaluated at FNAL, where 3 out of the 4 amplifiers, located in the output stages labeled 2, 3 and 4 in Figure~\ref{fig:ccddiagram}, were working as expected under operating conditions. Physically, the sensor has two serial registers, which are the last rows of pixels used to transfer the charge from the active area to the output stages. In this sensor, output stages 1 and 4, as well as 2 and 3, share a serial register. After installing the sensor in MOSKITA at the LHC, we found that half of the serial register next to output stage 2 presented problems in transferring charge with the usual operating voltages, as shown in Figure~\ref{fig:imagemoskita}. During data acquisition, each amplifier read out one quarter of the CCD (quadrant); hence, the only useful quadrants, named 1 and 2 from now on, are those read out through amplifiers in output stages 3 and 4, respectively.

\subsection{Data acquisition (DAQ) and operating conditions~\label{sec:daq}}
The data acquisition cycle for one image consists of two phases: cleaning and readout. During the cleaning phase, charge in the active area of the CCD is transferred and discarded through the output stages 1 and 2 for 10 minutes. Thus, the readout begins with a "clean" CCD. Immediately after cleaning, we start the pixel-by-pixel readout through the four amplifiers. Each amplifier reads 515 rows and 3200 columns of pixels, measuring $N_{\rm smp}$ times the charge in each pixel.

Of the 515 rows, 512 correspond to the active area, that is pixels within the physical array. The remaining 3 correspond to the vertical ``overscan'' area, which consists of pixels that extend beyond the physical array clocked towards the output stage. Of the 3200 columns, the first 8 correspond to the ``prescan'' area, which arises from 8 extra physical pixels in the serial register connecting to the output stage. The next 3072 columns correspond to the active area, followed by 20 columns of horizontal overscan. The remaining 100 columns correspond to  the horizontal ``reverse overscan'' area, that is pixels beyond the physical array clocked away from the output stage. The entire serial register is clocked to discard charge before each row of pixels in the active area is vertically transferred.

In each readout, data are stored as a two-dimensional FITS image with 515 rows and $(3200\times N_{\rm smp})$ columns, referred to as a ``raw'' image, to which a consecutive {\sout serial} ``image ID'' number is assigned. In this image, every $N_{\rm smp}$ consecutive pixels in the same row correspond to the $N_{\rm smp}$ measurements of the charge in the same pixel, and their values are in Analog-to-Digital Units (ADUs). It takes $\sim$3.2 (4.8)~hours to acquire one raw image with $N_{\rm smp}=169~(256)$~samples/pixel, resulting in a non-uniform exposure across pixels that increases linearly with the pixel readout time. Raw images are processed afterward for monitoring and analysis, as described in Section~\ref{sec:dataproc}. The pixel distributions of the processed images form a set of Gaussian peaks centered on integer multiples of the gain, a parameter that relates ADUs to the number of generated electron-hole pairs.

\begin{figure}[ht!]
    \centering
    \includegraphics[width=1\linewidth]{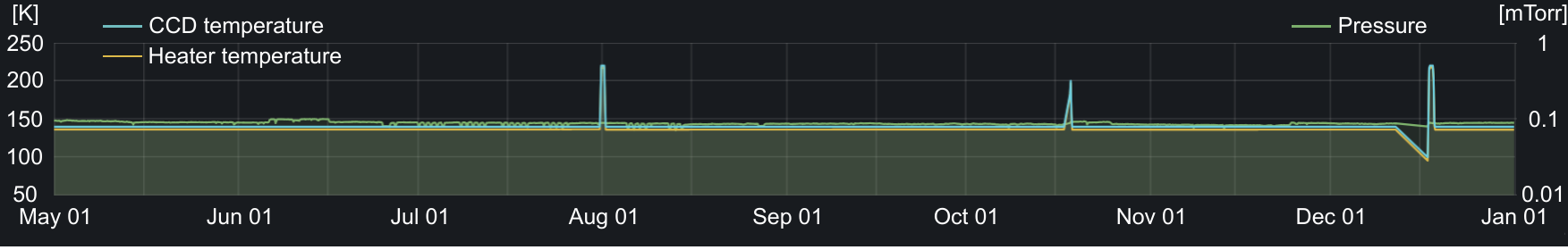}
    \caption{Pressure and temperature of the experimental setup over eight months within the DAQ period.}
    \label{fig:PTMoskita}
\end{figure}

MOSKITA at LHC has been taking data continuously since March 2024. We divide the DAQ period into ``runs'', each consisting of a set of images acquired under the same conditions. In this analysis, we use data from the first year of operations, whose information is summarized in Table~\ref{tab:data}.
In run 1, the sensor's operating voltages caused pixel saturation at $\sim$1400$e^-$ in quadrant 2. For subsequent runs, we changed the voltage applied to one of the gates in the sensor's output stage to achieve pixel saturation at typical levels, i.e. $\gtrsim8000e^-$ since a lower saturation value could impact event reconstruction at high energies. System thermal cycles to temperatures above 200K were performed after runs 2, 4 and 5. The operating conditions of the setup during each run were stable, maintaining a sensor temperature of 140K and a pressure below $1\times10^{-4}$~Torr; this is shown in Figure~\ref{fig:PTMoskita}.
\begin{table}[ht!]
\small
\centering
\begin{tblr}{c|cccc}
Run & Date start & Date end & No. of images & Conditions \\ \hline 
1 & Mar 15 & May 22 & 462 & Volt A; $N_{\rm smp}=169$\\
2 & May 22 & Jul 31 & 500 & Volt B; $N_{\rm smp}=169$\\
\hline[dotted]
3 & Aug 04 & Sep 14 & 293 & Volt B; $N_{\rm smp}=169$\\
4 & Sep 14 & Oct 17 & 159 & Volt B; $N_{\rm smp}=256$\\
\hline[dotted]
5 & Oct 19 & Dec 09 & 236 & Volt B; $N_{\rm smp}=256$\\
\hline[dotted]
6 & Dec 21 & Dec 31 & 50 & Volt B; $N_{\rm smp}=256$
\end{tblr}
\caption{Runs during DAQ period. Volt A/B refer to the sensor's operating voltages used in each run. Thermal cycles are indicated with dotted lines.}
\label{tab:data}
\end{table}

We continuously monitor the readout noise, gain and single-electron rate (SER) of each acquired image. These parameters are obtained from a three-Gaussian fit to the first three peaks of the pixel distributions of the processed images, corresponding to pixels with 0, 1 and 2$e^-$. Assuming the Gaussian fitted to the $ke^-$ peak has parameters $(a_k, \sigma_k, \mu_k)$ corresponding to its amplitude, standard deviation, and mean, respectively, we define the gain as $(\mu_2-\mu_0)/2$, the SER as $a_1/a_0$ and the readout noise as $\sigma_0$. 

\begin{figure}[ht!]
    \centering
    \includegraphics[width=\linewidth]{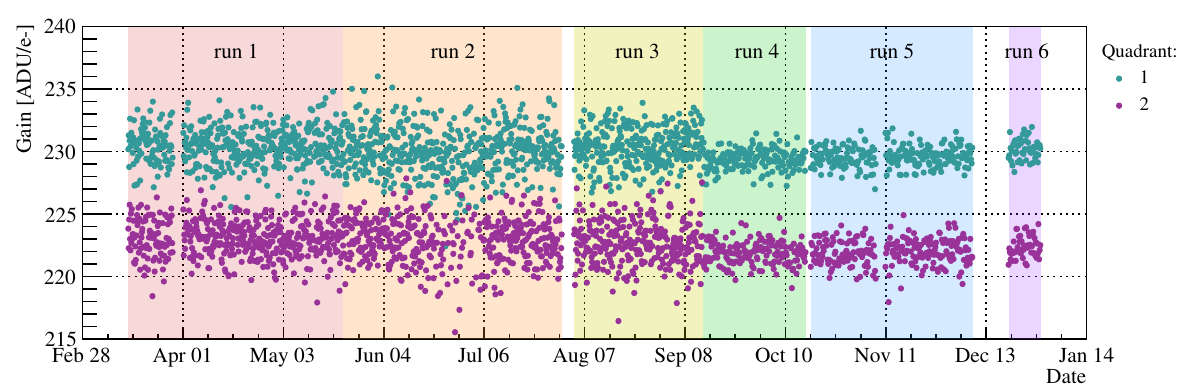}
    \caption{Gain evolution over time during the DAQ period. For reference, the time periods of the runs are highlighted.}
    \label{fig:gainmoni}
\end{figure}

\begin{figure}[ht!]
    \centering
    \includegraphics[width=\linewidth]{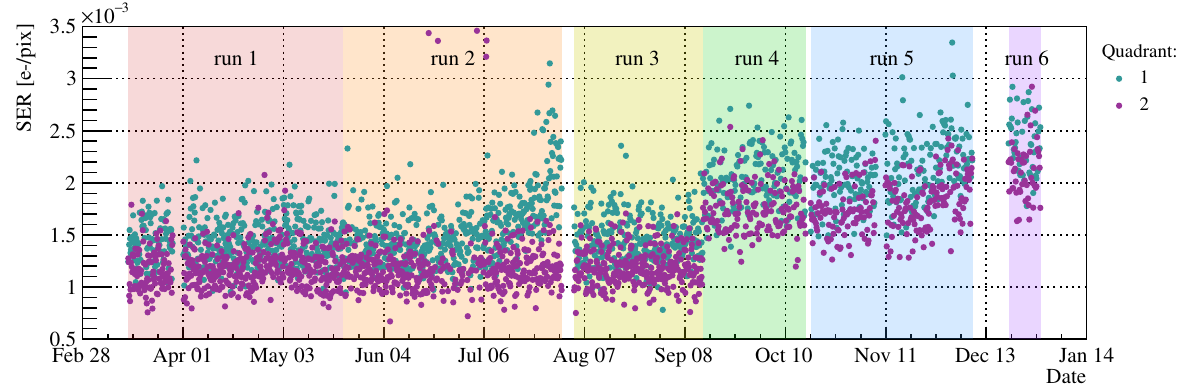}
    \caption{Single-electron rate evolution over time during the DAQ period. For reference, the time periods of the runs are highlighted.}
    \label{fig:sermoni}
\end{figure}

\begin{figure}[ht!]
    \centering
    \includegraphics[width=\linewidth]{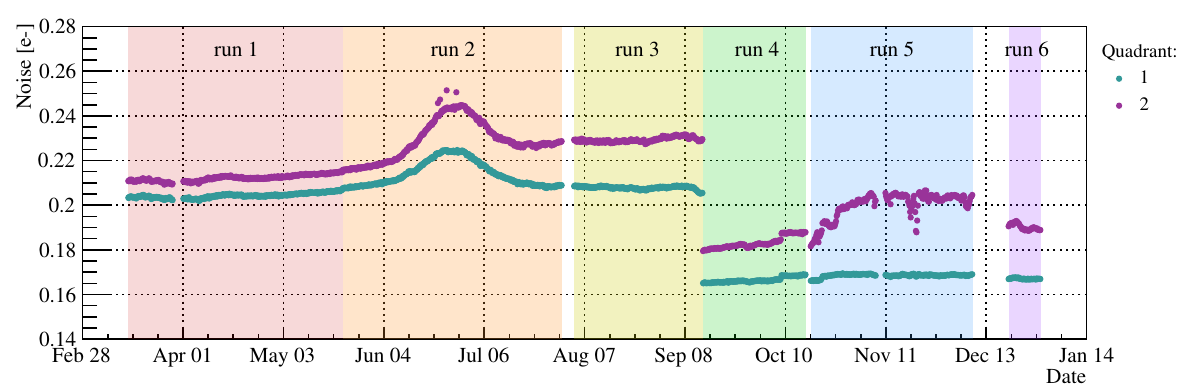}
    \caption{Readout noise evolution over time during the DAQ period. For reference, the time periods of the runs are highlighted.}
    \label{fig:noisemoni}
\end{figure}

Figures~\ref{fig:gainmoni}, \ref{fig:sermoni} and \ref{fig:noisemoni} show the evolution of these parameters in each working quadrant over time, with the SER and the noise scaled by the gain. The gain remains stable over time, with a smaller spread during runs 4, 5 and 6 due to improved fits resulting from the reduced noise at higher $N_{\rm smp}$. Similarly, the SER is stable overall in each run, with a slight increase observed toward the end of runs 2 and 5. As expected, the mean SER in runs 4, 5 and 6 is higher due to the longer image readout time associated with the increased $N_{\rm smp}$; however, the SER in run 6 is 17\% higher than in runs 4 and 5. In contrast, the readout noise shows significant variations over time exhibiting the highest variation during run 2, with 2.8\% for quadrant 1 and 3.5\% for quadrant 2. The mean value of each parameter, along with its standard deviation, for each quadrant in each run are shown in Table~\ref{tab:meandata}.
\begin{table}[ht!]
\small
\centering
\begin{tabular}{c | >{\centering}p{1.8cm} >{\centering}p{1.8cm} | >{\centering}p{1.8cm} >{\centering}p{1.8cm} | >{\centering}p{1.8cm} >{\centering\arraybackslash}p{1.8cm}}
\multirow{2}{*}{Run} & \multicolumn{2}{c|}{Gain [ADU/$e^-$]} & \multicolumn{2}{c|}{SER [$10^{-3}$~$e^-$/pix]} & \multicolumn{2}{c}{Noise [$e^-$]} \\ 
& Quadrant 1 & Quadrant 2 & Quadrant 1 & Quadrant 2 & Quadrant 1 & Quadrant 2 \\
\hline
1 & $230.4\pm1.5$ & $223.1\pm1.4$ & $1.4\pm0.2$ & $1.2\pm0.2$ & $0.204\pm0.001$ & $0.212\pm0.001$ \\
2 & $230.0\pm1.9$ & $222.9\pm1.7$ & $1.5\pm0.3$ & $1.2\pm0.2$ & $0.214\pm0.006$ & $0.227\pm0.008$ \\
3 & $230.4\pm1.5$ & $222.7\pm1.7$ & $1.4\pm0.2$ & $1.2\pm0.2$ & $0.208\pm0.001$ & $0.229\pm0.001$ \\
4 & $229.4\pm0.8$ & $221.9\pm0.9$ & $2.0\pm0.3$ & $1.7\pm0.2$ & $0.166\pm0.001$ & $0.183\pm0.003$ \\
5 & $229.6\pm0.8$ & $222.0\pm1.0$ & $2.1\pm0.3$ & $1.8\pm0.2$ & $0.169\pm0.001$ & $0.200\pm0.006$\\
6 & $230.2\pm0.8$ & $222.4\pm0.8$ & $2.4\pm0.3$ & $2.1\pm0.3$ & $0.167\pm0.001$ & $0.190\pm0.001$
\end{tabular}
\caption{Gain, single-electron rate and readout noise mean values for each quadrant in each run.}
\label{tab:meandata}
\end{table}

\subsubsection{The LHC beam during DAQ}
DAQ ocurred during Run~3 of the LHC. In its 2024 run, the LHC conducted a proton run from March to October, followed by a lead-ion run that concluded in November.

Proton-proton collisions at an energy of 13.6~TeV began on March 14. After achieving stability, the beam intensity steadily increased, reaching its maximum at the end of April with a recorded instantaneous luminosity of $\sim$$2.12\times10^{34}$~cm$^{-2}$~s$^{-1}$. At full intensity, the LHC beam consisted of 2352 bunches separated by 25~ns, each containing $1.6\times10^{11}$~protons. The beam operates in periods of $\sim$10 hours, during which the instantaneous luminosity maintains its initial peak for $\sim$4~hours before decaying exponentially until the beam is dumped. The proton run concluded on October 16, delivering a final integrated luminosity of 124.4~fb$^{-1}$ to the CMS experiment.

After the proton run, a 6-day period of proton-proton collisions at 5.36~TeV took place for calibration for the lead-ion run. Pb-Pb collisions at 5.36~TeV began on November 6. Full intensity was achieved on November 10, with 1240 bunches per beam separated by 50~ns. During this run, the beam operated in periods of $\sim$6~hours, maintaining its initial luminosity peak for about one hour before exponentially decaying until being dumped. The lead-ion run concluded on November 23, delivering a final integrated luminosity of 1.9~nb$^{-1}$ to the CMS experiment.

\begin{figure}[ht!]
\centering
\includegraphics[width=\textwidth]{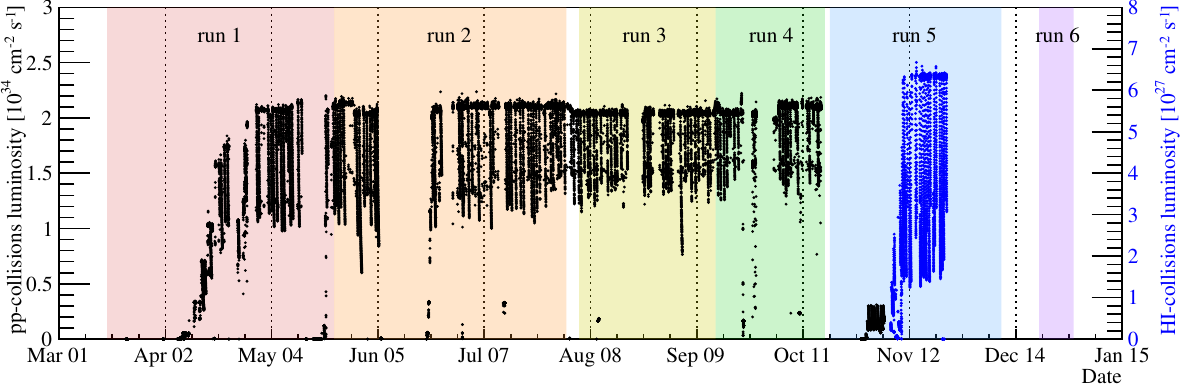}
\caption{Instantaneous luminosity of the LHC beam recorded by the CMS detector during the 2024 run for proton-proton (black) and lead-lead (blue) collisions. For reference, the time periods of MOSKITA DAQ runs are highlighted.}
\label{fig:luminosity}
\end{figure}

Since the DAQ is not synchronized with the beam periods, we associate to each acquired image the integral of the instantaneous luminosity over its readout time. Figure~\ref{fig:luminosity} shows the LHC beam instantaneous luminosity recorded by the CMS detector during the 2024 run.

\section{Data processing~\label{sec:dataproc}}
We perform a two-step image processing. The first step is applied to each raw image, where we build its corresponding ``processed'' image by assigning the average of its $N_{\rm smp}$ measurements, excluding the first one which is noisier, to each pixel. As a result, the processed image has $N_{\rm smp}$ times fewer columns compared to the raw image. Additionally, the processed image includes a row-by-row subtraction of the horizontal baseline, computed as the median of the horizontal reverse overscan pixels in the row, which are considered to have zero charge. The pixel values in the processed images are in Analog-to-Digital Units (ADUs).

A second processing step is applied to each processed image, where we first perform a column-by-column subtraction of the vertical baseline, computed as the median of all pixels in the column, and then we build its corresponding ``calibrated'' image by dividing each pixel value by the absolute gain per quadrant, computed as described in Section~\ref{sec:calibration}. Pixel values in the calibrated images are in units of electrons.

\subsection{Energy calibration~\label{sec:calibration}} 
\begin{figure}[ht!]
\centering
\includegraphics[width=1\textwidth]{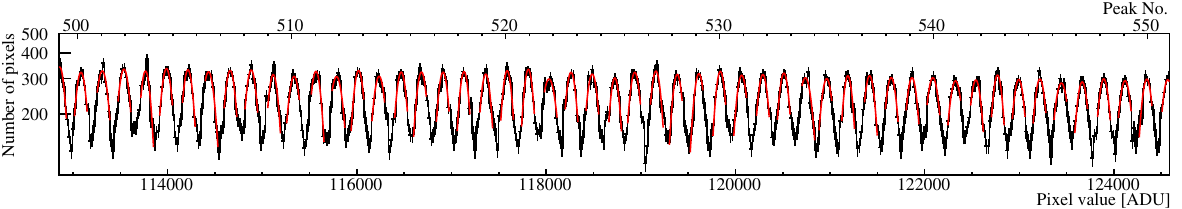}
\caption{Pixel distribution corresponding to quadrant 1 in run 2 in ADUs. The peak number is equivalent to the number of collected charge carriers.}
\label{fig:epeaksforgain}
\end{figure}

For each run and quadrant, we sum the pixel distributions of all processed images and perform a Gaussian fit to each of the first 850 peaks. Figure~\ref{fig:epeaksforgain} shows the fitted pixel distribution corresponding to quadrant 1 in run 2, between peak numbers 500 and 550. We plot the mean value of each of those fits against its corresponding peak number and perform a linear fit; the slope extracted from this fit corresponds to the average gain for each quadrant in each run.  We find this gain to be essentially constant over time. These values are used for calibration and are shown in Table~\ref{tab:gaincal}. To estimate the non-linearity (N.L.) in the energy calibration, we compare the mean of each Gaussian fit, divided by the average gain for each quadrant in each run, to its corresponding peak number and extract the highest percentage variation. These values represent upper limits on the energy calibration non-linearity, which are below 1.9\% for quadrant 1 and 4\% for quadrant 2 in all runs; see Table~\ref{tab:gaincal}. The gain values obtained in Section~\ref{sec:daq} are consistent with this method when taking into account the non-linearity.

\begin{table}[ht!]
\small
\centering
\begin{tabular}{c|cccccccccccc}
\multirow{2}{*}{Quadrant} & \multicolumn{2}{c}{Run 1} & \multicolumn{2}{c}{Run 2} & \multicolumn{2}{c}{Run 3} & \multicolumn{2}{c}{Run 4} & \multicolumn{2}{c}{Run 5} & \multicolumn{2}{c}{Run 6}\\ 
& Gain & N.L. & Gain & N.L. & Gain & N.L. & Gain & N.L. & Gain & N.L. & Gain & N.L.\\
\hline
1 & $226.2$ & 1.8\% & $226.3$ & 1.7\% & $226.3$ & 1.9\% & $226.3$ & 1.3\% & $226.4$ & 1.5\% & $226.8$ & 1.7\%\\
2 & $214.7$ & 4\% & $214.8$ & 4\% & $214.8$ & 4\% & $214.7$ & 3.4\% & $214.8$ & 3.4\% & $215.1$ & 3.6\%\\
\end{tabular}
\caption{Average gain in ADU/$e^-$ and the highest percentage variation in energy calibration due to non-linearity for each quadrant in each run.}
\label{tab:gaincal}
\end{table}

\section{Event reconstruction}
From the calibrated images, we reconstruct events. We define an event as a group of neighboring pixels with charges exceeding the 1$e^-$ detection threshold, computed per image as described in Section~\ref{sec:detthr}. The total charge of an event is computed as the sum of the integer charges of its constituent pixels.

\subsection{Detection threshold}~\label{sec:detthr}
\begin{figure}[ht!]
\centering
\includegraphics[width=1\textwidth]{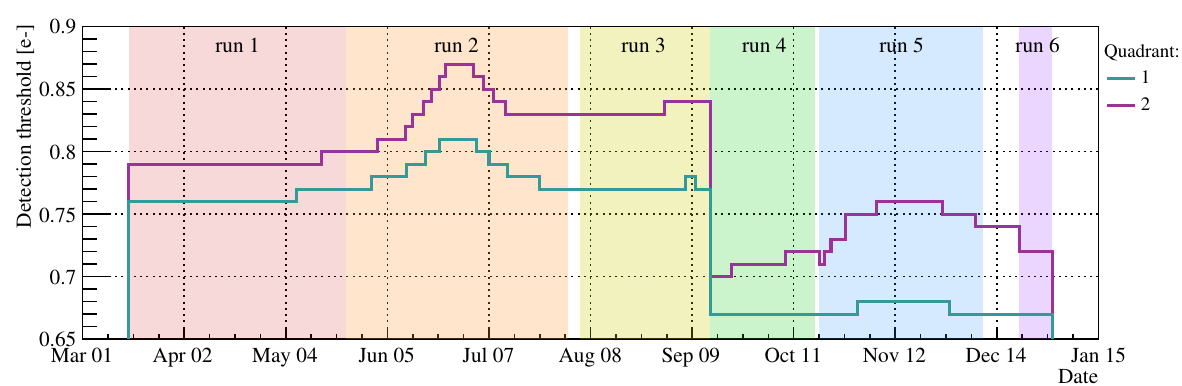}
\caption{Optimal 1$e^-$ detection threshold per image for quadrant 1 (cyan) and 2 (purple) in each run.}
\label{fig:detthr}
\end{figure}

The minimum amount of charge that a pixel must have to be considered as non-empty, i.e. the 1$e^-$ detection threshold, strongly depends on the noise and the single-electron rate of the image. During the data acquisition period, the SER remained stable, but the noise exhibited some variations, as discussed in Section~\ref{sec:daq}. To select the 1$e^-$ detection threshold accounting for these variations, we proceed as follows. We model the 0$e^-$ and 1$e^-$ pixel distributions of each image as two Gaussians centered at 0 and 1, respectively, with the same standard deviation determined by the image noise value in units of electrons. The amplitude ratio of the 1$e^-$ Gaussian to the 0$e^-$ Gaussian equals the image SER in $e^-$/pix. We compute the F1 score for the 1$e^-$ signal as a function of the threshold value. We define the optimal threshold for each image as the one that maximizes the F1 score. Figure~\ref{fig:detthr} shows the optimal detection threshold for each image of each quadrant over the DAQ period.

\section{Event selection} \label{sec:evtsel}
We use masks, created from the calibrated images, to exclude events from known sources. The masking algorithm, described in Section~\ref{sec:masks}, is based on the approach used in dark matter searches~\cite{SENSEI2020, SENSEIsnolab2025}. Low-energy events, with charge $\leq20e^-$, are rejected if any of their constituent pixels is masked by any of the masks, while high-energy events, with charge $>20e^-$, are rejected if their energy-weighted pixel centroid is masked by the border, hot-column or serial-register-event masks.

We apply an additional selection cut to the low-energy events, beyond the masking-rejection criterion, to exclude single-row events. This cut targets events likely happening near the sensor surface, primarily causing single-pixel events, and artifacts from serial-register hits or charge-transfer inefficiencies in the output stage, which manifest as one-row multiple-pixel events.

The impact of the masks and the selection cut on the detection of low-energy events is incorporated as an efficiency, calculated for each energy bin. This efficiency is computed from simulated images, as detailed in Section~\ref{sec:deteff}, and accounts for various effects that affect the event reconstruction, including:
\begin{enumerate}
\item Diffusion: Events may split into multiple clusters when diffusing, or part of their charge may fall into masked regions, resulting in only a fraction of the true deposited charge being reconstructed.
\item Noise: Per-pixel fluctuations can result in events being reconstructed with a charge differing from the actual deposited charge.
\item Single-electron rate: Multi-electron events can form from pile-up, and events overlapping with 1$e^-$ events from the SER may be reconstructed with a charge higher than originally deposited.
\end{enumerate}

\subsection{Masking algorithm}~\label{sec:masks}
We developed and optimized a masking algorithm using only images from run~1, in order to avoid any bias, and then apply it to the full dataset for the final analysis. A brief discussion of the masks used in this analysis is presented here:

\textbf{Border mask:} This mask removes events near the image borders, as their measured energy may only be a fraction of the total energy deposited due to their proximity to the edge. Additionally, low-energy events near the borders can originate from depositions occurring outside the active area of the sensor. We mask 10 pixels from the top, bottom and left edges in the images.

\begin{figure}[ht!]
    \centering
    \includegraphics[width=0.75\linewidth]{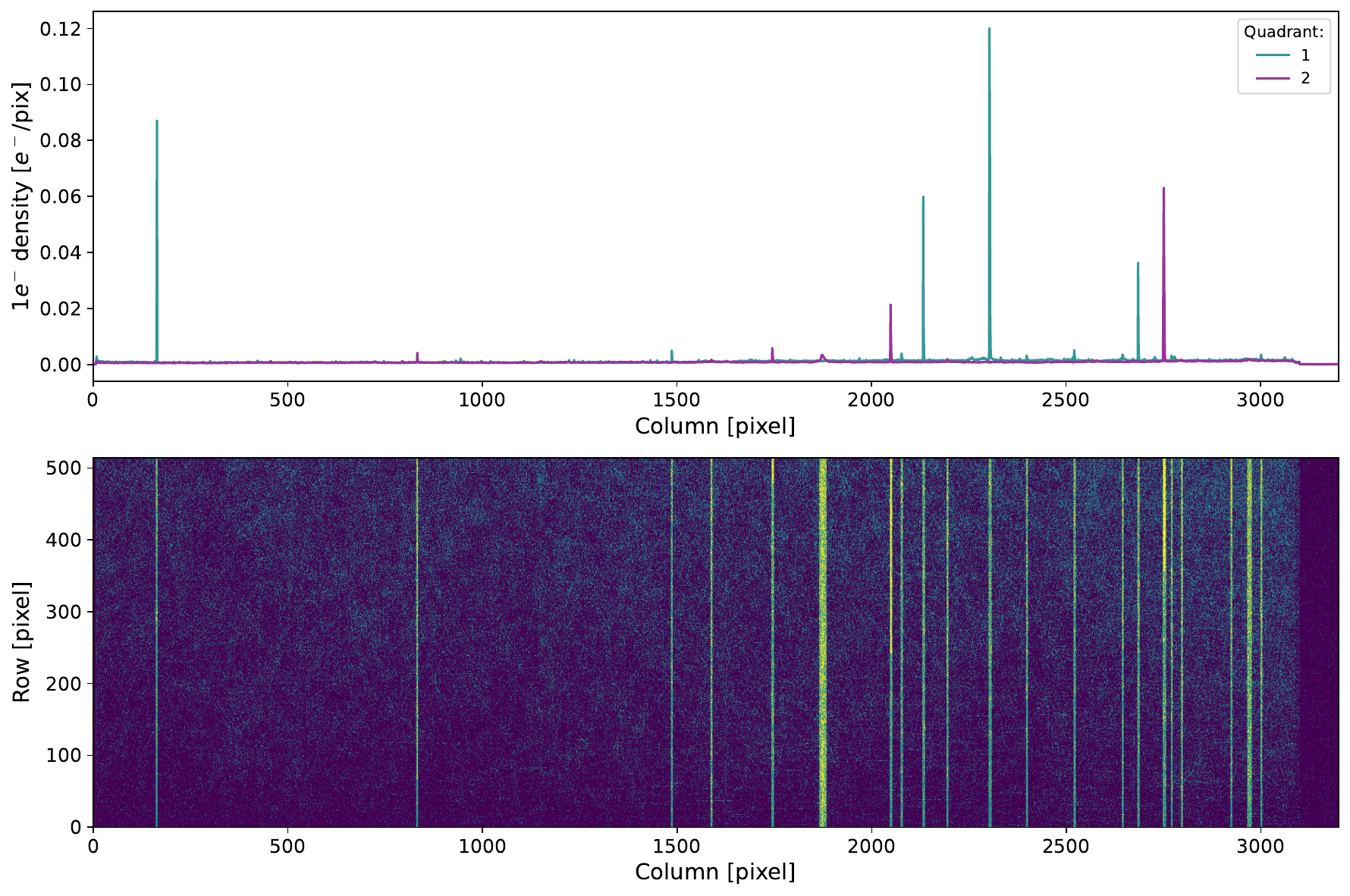}
    \caption{Top: 1$e^-$ event rate per column for quadrants 1 (cyan) and 2 (purple). Bottom: Hot-column mask for run 2.}
    \label{fig:hotcolsmask}
\end{figure}

\textbf{Hot-column mask:} This mask removes columns with an excess of 1$e^-$ events produced by defects in the sensor's active area. For each run, a single mask is generated by combining the hot-column mask of the two useful quadrants. To create the mask of each quadrant, the 1$e^-$ event rate for each column is computed using all processed images from that run, after removing all events with a charge greater than 20$e^-$ and a halo of 5 pixels around them. A smoothing filter is applied to the rate, the baseline is removed, and columns with a rate exceeding 5 median absolute deviations, along with their neighboring columns, are masked. As an example, we show in Fig.~\ref{fig:hotcolsmask} the 1$e^-$ event rate per column for each quadrant together with the hot-column mask for run 2.

\begin{figure}[ht!]
    \centering
    \includegraphics[width=0.78\linewidth]{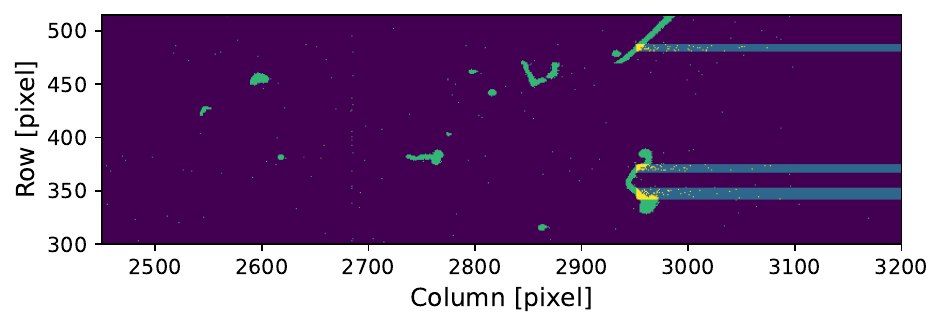}
    \caption{Transfer-gate-trap mask for an image of quadrant 1. Events containing a high-energy pixel in column 2952 show a deferred charge tail. The mask is shown superimposed on the image, and non-empty pixels inside the masked region are highlighted in yellow.}
    \label{fig:Tgate_trap}
\end{figure}

\textbf{Transfer-gate-trap mask:} We identified a horizontal ``tail'' of low-energy events following high-energy events with at least one pixel in column 2952 of quadrant 1. We attribute the origin of the events forming the tail to charge emission from a charge trap in the sensor's transfer gate in column 2952. This gate enables charge transfer from the last pixel in that column to the serial register; therefore, a trap in this gate can capture charge and release it later into the serial register. To reject events originating from this trap, we mask the pixels with charge greater than 1$e^-$ in column 2952, along with all pixels to their right in the same row, in all images from quadrant 1. Fig.~\ref{fig:Tgate_trap} shows this mask for an image of quadrant 1.

\begin{figure}[ht!]
    \centering
    \includegraphics[width=0.49\linewidth]{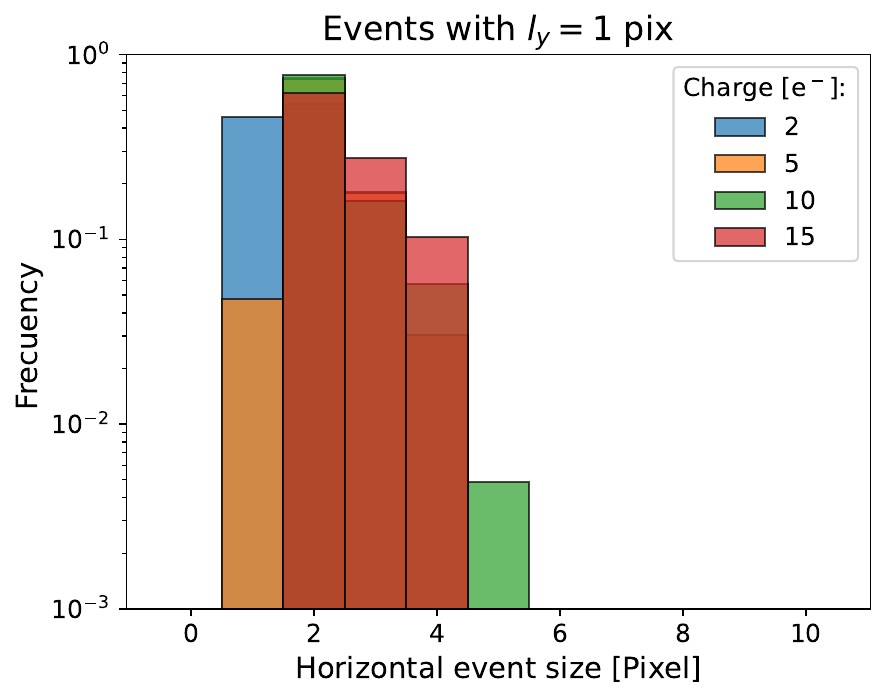}
    \includegraphics[width=0.49\linewidth]{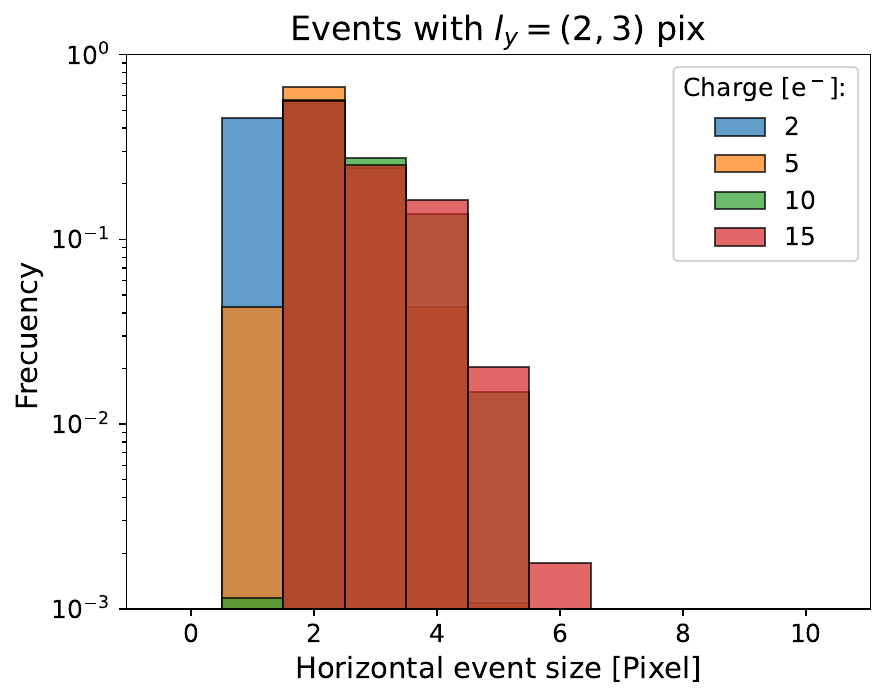}
    \caption{Horizontal size distributions of simulated events with different charge values and vertical sizes: $l_y=1$~pixel (left) and $l_y=2$~or~3 pixels (right). For the plots, $10^6$ events were simulated per charge value.}
    \label{fig:sizespread_dif}
\end{figure}

\textbf{Serial-register-event mask:} This mask, created for each image, removes primarily horizontal, single-row events, which arise from charge depositions either directly within the serial register or diffusing into it from the surrounding inactive silicon. After applying the hot-column and the transfer-gate-trap masks, we flag events as serial-register events based on two criteria: 1) their size and 2) the distance between events. For the size-based criteria, events are flagged if they meet either of the following conditions: a) $l_y=1$~pixel and $l_x>4$~pixels, or b) $l_y=2$~or~3 pixels and $l_x>7$~pixels, where $l_x$ and $l_y$ are the event sizes in the horizontal (row) and vertical (column) directions, respectively. These conditions are based on the event sizes of point-like charge depositions occurring on the backside of the sensor, computed from simulations using the diffusion model described in Ref.~{\cite{SENSEI2020}}. Figure~\ref{fig:sizespread_dif} shows the horizontal size distributions of simulated events with $l_y=1$~pixel (left) and $l_y=2$~or~3 pixels (right), with varying charge values.

\begin{figure}[ht!]
    \centering
    \includegraphics[width=0.49\linewidth]{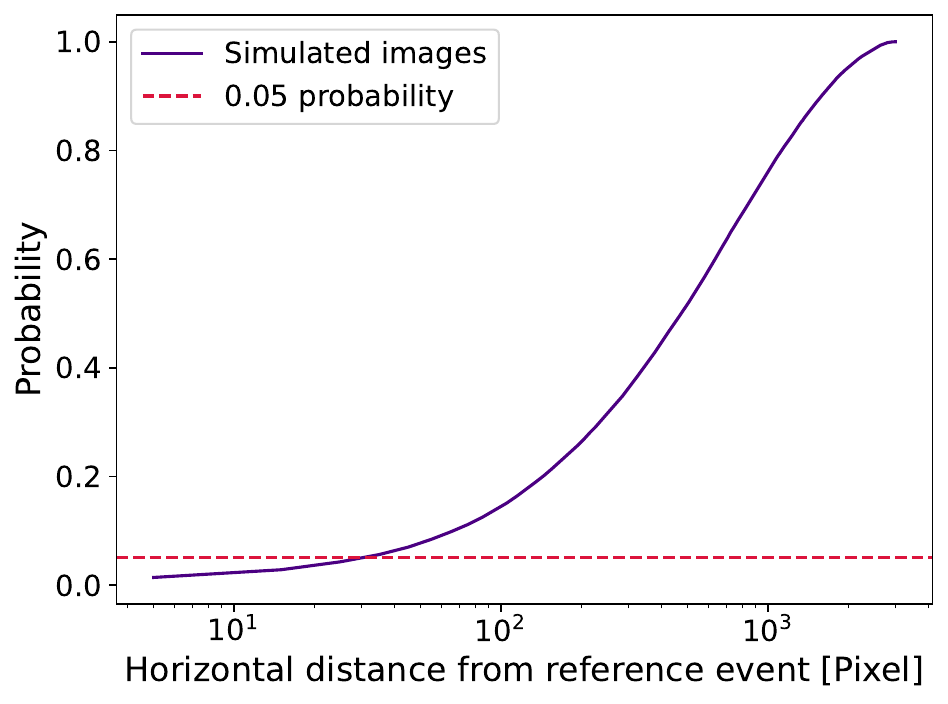}
    \includegraphics[width=0.49\linewidth]{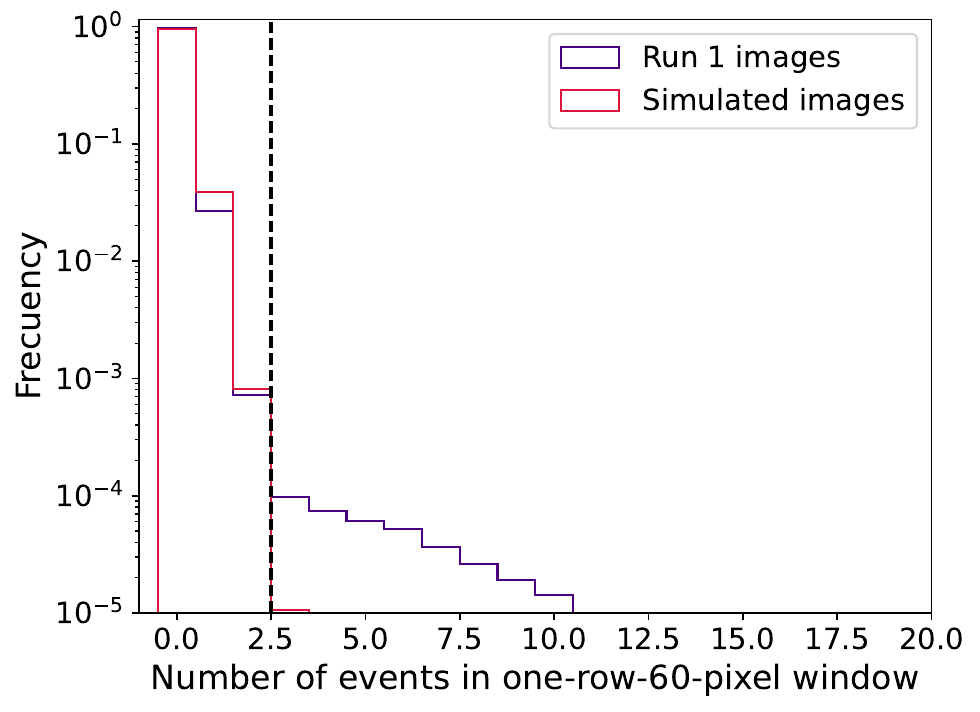}
    \caption{Left: Probability of finding an event as a function of the horizontal distance from a reference event in simulated images (blue). The 30-pixel criterion for non-single-pixel events corresponds to a $\sim$5\% probability (dashed red). Right: Number of single-pixel events in 60-pixel windows along one row for simulated (red) and run 1 (blue) images. Events to the right of the dashed black line are flagged as serial-register events.}
    \label{fig:prob_spatialdist}
\end{figure}

The distance-based criteria for flagging an event with $l_y\leq3$~pixels as a serial-register event are as follows: 1) non-single-pixel events must have a shape consisting of a single row with $l_x>1$~pixel and immediate upper and lower rows with $l_x\leq1$~pixel, and be within a horizontal distance of 30 pixels from another event with $l_y\leq3$~pixels; 2) single-pixel events require three or more of such events in the same row within a 60-pixel window. These criteria are based on the distance between events in simulated images containing uniformly distributed single-pixel events, with their total number matching the observed events with $l_y\leq3$~pixels in run 1 images. Figure~\ref{fig:prob_spatialdist} (left) shows the probability of finding at least one event at a given horizontal distance from a randomly chosen reference event, computed from the simulated images, with the 30-pixel criterion corresponding to a $\sim$5\% probability. Figure~\ref{fig:prob_spatialdist} (right) shows the number of single-pixel events in one-row 60-pixel windows for the simulated and run 1 images. Once an event is flagged as a serial-register event, we mask the entire row(s) associated with it.

\textbf{Crosstalk mask:} Electronic crosstalk between amplifiers can cause a small signal to leak from one channel with a strong signal to another. To remove events caused by crosstalk, we create a mask for each image that flags a pixel if it was read at the same time as another pixel with a charge higher than 1000$e^-$.

\textbf{Bleeding-zone mask:} This mask, created for each image, removes spurious events from horizontal and vertical charge-transfer inefficiencies. For each pixel with a charge exceeding 20$e^-$, we mask 100 pixels to the right and 30 pixels above. These values were determined by analyzing the SER in run 1 images after applying the hot-column, serial-register-event, transfer-gate-trap, and cross-talk masks. The chosen values correspond to the number of masked pixels beyond which the SER remained constant.

\textbf{Halo mask:} High-energy events correlate with an increased rate of low-energy events in nearby pixels, associated with secondary radiation. This mask, created for each image, flags all pixels within a 25-pixel radius of those with charges exceeding 20$e^-$. The radius was determined similarly to the parameters in the bleeding-zone mask, based on the radius at which the SER stabilized after applying the hot-column, serial-register-event, transfer-gate-trap, cross-talk and bleeding-zone masks.

\begin{figure}[ht!]
    \centering
    \includegraphics[width=\linewidth]{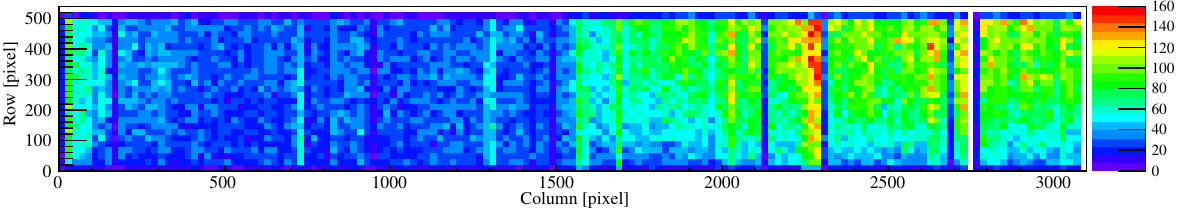}
    \includegraphics[width=\linewidth]{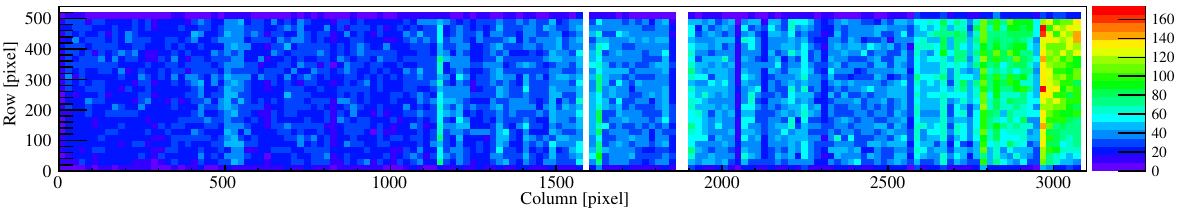}
    \caption{2D histograms showing the spatial distribution of 1$e^-$ events from all run 1 images after applying the border, hot-column, serial-register-event, transfer-gate-trap, cross-talk, bleeding-zone and halo masks for quadrant 1 (top) and quadrant 2 (bottom). Each bin represents 20~pixels $\times$ 20~pixels.}
    \label{fig:hotzones}
\end{figure}

\textbf{Hot-zone mask:} After applying all the previously discussed masks to a set of images, the combined event distribution is expected to be spatially uniform. However, when plotting the spatial distribution of the unmasked 1$e^-$ events per quadrant for all images in run 1, we identify regions in the sensor with an excess of events, referred to as ``hot zones''. Figure~\ref{fig:hotzones} shows the combined spatial distribution of unmasked 1$e^-$ events from all run 1 images per quadrant, clearly highlighting the hot zones. This mask removes these zones.

\subsection{Detection efficiency for low-energy events}~\label{sec:deteff}
We compute the detection efficiency for events with a charge between 2 and 20$e^-$ from simulated images. To account for variations in noise and SER during the data-taking period, we construct each simulated image using the parameters and corresponding mask of a randomly selected image from all runs.

Each simulated image contains spatially uniform 1$e^-$ events, with a total number matching the SER of the selected image, and a single multi-electron event with a charge between 2 and 21$e^-$, simulated below unmasked pixels. The deposition depth of the multi-electron event is randomly selected from a uniform distribution along the sensor depth, and its charge is diffused toward the surface using the model described in Ref.~{\cite{SENSEI2020}}. To account for readout noise, the charge in each pixel is adjusted based on the noise of the selected image. Events are then reconstructed using the energy threshold corresponding to the selected image, and the masking-rejection criterion and the selection cut are applied.

From a set of 50,000 simulated images, the detection efficiency for each low-energy bin is calculated as the ratio of the total number of simulated multi-electron events with that energy to the number of reconstructed and non-rejected multi-electron events with the same energy.

\section{Event rates}
For the period of proton-proton collisions, we include images from runs 1 to 4 and run 6, resulting in a total exposure of 116.20~g-day, combining data from both quadrants, and a luminosity of 113.3 fb$^{-1}$. For the period of Pb-Pb collisions, we include images from runs 5 and 6, yielding a total exposure of 31.85~g-day, combining data from both quadrants, and a luminosity of 1.54~nb$^{-1}$. Note that images from run 6 are included in both periods, as they correspond to a time without collisions. Results for the low-energy region ($\leq20e^-$) are presented in Section~\ref{sec:lowEresults}, and those for the high-energy region ($>20e^-$) are discussed in Section~\ref{sec:highEresults}.

\subsection{Low-energy region} \label{sec:lowEresults}
The total post-masks exposure, $\varepsilon_{tot}$ (ALL), is calculated from the unmasked regions in the images for each collision period after applying the full set of masks, yielding 39.32~g-day for proton-proton collisions and 9.45~g-day for Pb-Pb collisions. The post-masks exposure for images with zero luminosity, $\varepsilon_{tot}$ (L=0), is 18.24~g-day for the proton-proton collision period and 7.65~g-day for the Pb-Pb collision period.

Results by energy bin are shown in Table~\ref{tab:lowEpp} for proton-proton collisions and Table~\ref{tab:lowEii} for Pb-Pb collisions. For each period, we count the number of observed events in each bin, $N$, combining data from both quadrants, after applying the low-energy event selection described in Section~\ref{sec:evtsel}. These numbers are reported in Tables~\ref{tab:lowEpp} and \ref{tab:lowEii} as total events (ALL) and as events occurring in images with zero luminosity (L=0). Detection efficiencies for each bin, $\mathit{Eff}$, computed as described in Section~\ref{sec:deteff}, are also shown in Tables~\ref{tab:lowEpp} and \ref{tab:lowEii}. The effective exposures for each bin, computed as $\epsilon=\mathit{Eff} \times \varepsilon_{tot}$, are included in these tables. Note that the event rate, $R$, can be computed as $R=N/\epsilon$.

To investigate potential correlations between the observed events and the LHC luminosity, we perform a likelihood analysis testing the background-only hypothesis ($\mu$=0), as described in Section~\ref{sec:likelihood}, for each energy bin and collision period. The corresponding p-values, $p_0\equiv p(\mu=0)$, for each bin are reported in Tables~\ref{tab:lowEpp} and \ref{tab:lowEii} for the proton-proton and the Pb-Pb collision periods, respectively.

For the proton-proton collision period, the results are consistent with the background-only hypothesis. In the case of the Pb-Pb collision period, the analysis for the 7$e^-$ bin yields the lowest p-value, suggesting a possible deviation from the background-only hypothesis. However, this deviation may still be attributed to a statistical fluctuation. According to the "look-elsewhere effect", the probability of observing at least one such fluctuation with p-value $p^*$ in $n$ bins is $\mathcal{P}=1-(1-p^*)^n$, which in our case yields $\mathcal{P}=0.29$.

The 95\% C.L. upper limits on the number of signal events for each energy bin, computed as described in Section~\ref{sec:likelihood}, are also listed in Tables~\ref{tab:lowEpp} and \ref{tab:lowEii} for both collision periods. Considering the results in the 4$e^-$-7$e^-$ energy range, the region used for the mCP search, the corresponding p-values for testing the null hypothesis are $p_0=0.096$ for the proton-proton collision period and $p_0=0.11$ for the Pb-Pb collision period. Combining these p-values, we get an overall $p_0=0.059$, which approaches statistical significance (2$\sigma$). Unaccounted low-energy backgrounds that scale with beam luminosity may contribute to the observed result. One such potential background is low-energy neutrons arriving at the detector from secondary processes from high-energy collisions.

\begin{table}[ht!]
\centering
\footnotesize
\begin{tabular}{|c|>{\centering\arraybackslash}p{1cm}>{\centering\arraybackslash}p{1cm}|>{\centering\arraybackslash}p{1.2cm}>{\centering\arraybackslash}p{1.2cm}|c|c|c|}
\hline
\multirow{2}{*}{\textbf{Bin [$\mathbf{e^-}$]}} & \multicolumn{2}{c|}{\textbf{Observed events}} & \multicolumn{2}{c|}{\textbf{Eff. exposure [g-day]}} & \multirow{2}{*}{\textbf{Efficiency}} & \textbf{P-value} & \textbf{Upper limit 95\% C.L.} \\
& \textbf{ALL} & \textbf{L=0} & \textbf{ALL} & \textbf{L=0} &  & $\mu=0$ & \textbf{[events]} \\ \hline
2 & 327 & 161 & 11.40 & 5.29 & 0.29 & 0.31 & 5.88  \\ \hline 
3 & 17 & 11 & 19.27 & 8.94 & 0.49 &  0.20 & 1.41 \\ \hline
4 & 1 & 0 & 23.59 & 10.94 & 0.60 & 0.12 & 5.20 \\ \hline
5 & 2 & 0 & 26.74 & 12.40 & 0.68 & 0.37 &  3.12 \\ \hline
6 & 0 & 0 & 27.92 & 12.95 & 0.71 & 1 & 2.45 \\ \hline
7 & 1 & 0 & 29.10 & 13.50 & 0.74 & 0.23 &  4.15 \\ \hline
8 & 2 & 1 & 29.49 & 13.68 & 0.75 & 0.35 &  1.41 \\ \hline
9 & 0 & 0 & 30.67 & 14.23 & 0.78 & 1 & 1.81\\ \hline
10 & 1 & 0 & 31.85 & 14.77 & 0.81 & 0.53&  3.73 \\ \hline
11-20 & 9 & 3 & 33.42 & 15.50 & 0.85 & 0.41 & 8.17 \\ \hline
\end{tabular}
\caption{Results for the period of proton-proton collisions. We expect 271 2$e^-$ events and 1 3$e^-$ event from the pile-up of 1$e^-$ events, computed considering the SER of the images for the period.}
\label{tab:lowEpp}
\end{table}

\begin{table}[ht!]
\centering
\footnotesize
\begin{tabular}{|c|>{\centering\arraybackslash}p{1cm}>{\centering\arraybackslash}p{1cm}|>{\centering\arraybackslash}p{1.2cm}>{\centering\arraybackslash}p{1.2cm}|c|c|c|}
\hline
\multirow{2}{*}{\textbf{Bin [$\mathbf{e^-}$]}} & \multicolumn{2}{c|}{\textbf{Observed events}} & \multicolumn{2}{c|}{\textbf{Eff. exposure [g-day]}} & \multirow{2}{*}{\textbf{Efficiency}} & \textbf{P-value} & \textbf{Upper limit 95\% C.L.} \\
& \textbf{ALL} & \textbf{L=0} & \textbf{ALL} & \textbf{L=0} &  & $\mu=0$ & \textbf{[events]} \\ \hline
2 & 85 & 72 & 2.74 & 2.22 & 0.29 & 0.52  & 5.80 \\ \hline
3 & 6 & 5 & 4.63 & 3.75 & 0.49 & 0.43    & 2.05 \\ \hline
4 & 0 & 0 & 5.67 & 4.59 & 0.60 & 1 & 1.14 \\ \hline
5 & 2 & 2 & 6.43 & 5.20 & 0.68 & 0.57 & 2.02 \\ \hline
6 & 0 & 0 & 6.71 & 5.43 & 0.71 & 1 & 1.50\\ \hline
7 & 2 & 0 & 6.99 & 5.66 & 0.74 & 0.017 &  5.74 \\ \hline
8 & 0 & 0 & 7.09 & 5.74 & 0.75 & 1 & 1.74 \\ \hline
9 & 0 & 0 & 7.37 & 5.97 & 0.78 & 1 & 1.46 \\ \hline
10 & 0 & 0 & 7.65 & 6.20 & 0.81 & 1 & 2.30 \\ \hline
11-20 & 6 & 4 & 8.03 & 6.50 & 0.85 & 0.71 & 4.87 \\ \hline
\end{tabular}
\caption{Results for the period of Pb-Pb collisions. We expect 54 2$e^-$ events from the pile-up of 1$e^-$ events, computed considering the SER of the images for the period.}
\label{tab:lowEii}
\end{table}

\subsection{High-energy region} \label{sec:highEresults}
\begin{figure}[ht!]
\centering
\includegraphics[width=0.49\linewidth]{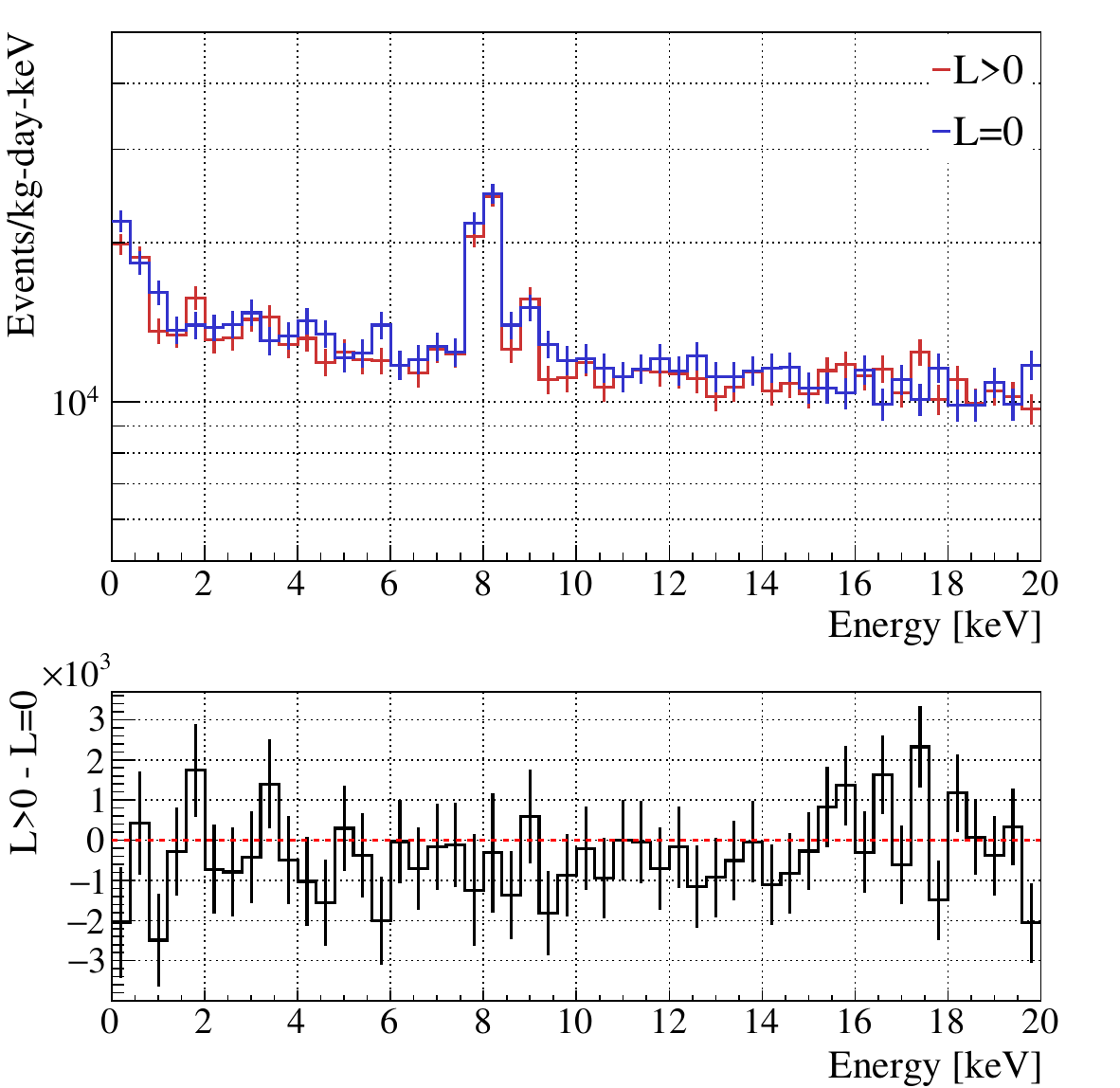}\hfill
\includegraphics[width=0.49\linewidth]{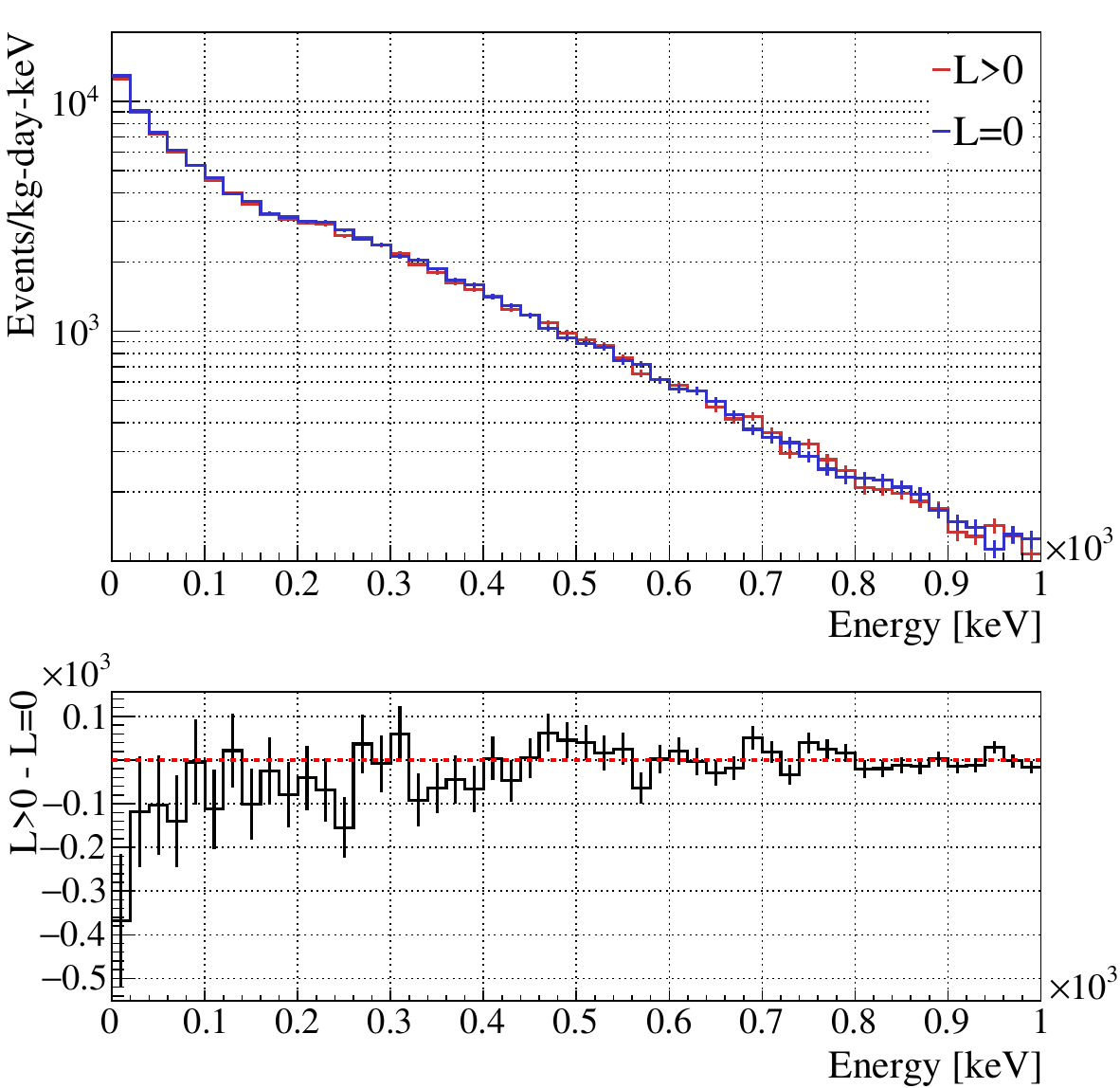}
\caption{Top: High-energy event rates for the proton-proton collision period, for images with L=0 (blue) and L$>$0 (red), shown from 0 to 20~keV (left) and from 0 to 1000~keV (right). Bottom: Difference in event rates between images with and without luminosity (black), i.e. L$>$0 - L=0, for the same energy ranges. A higher event rate below $\sim$450~keV for L=0 is observed.}
\label{fig:highEspecpp}
\end{figure}

\begin{figure}[ht!]
\centering
\includegraphics[width=0.49\linewidth]{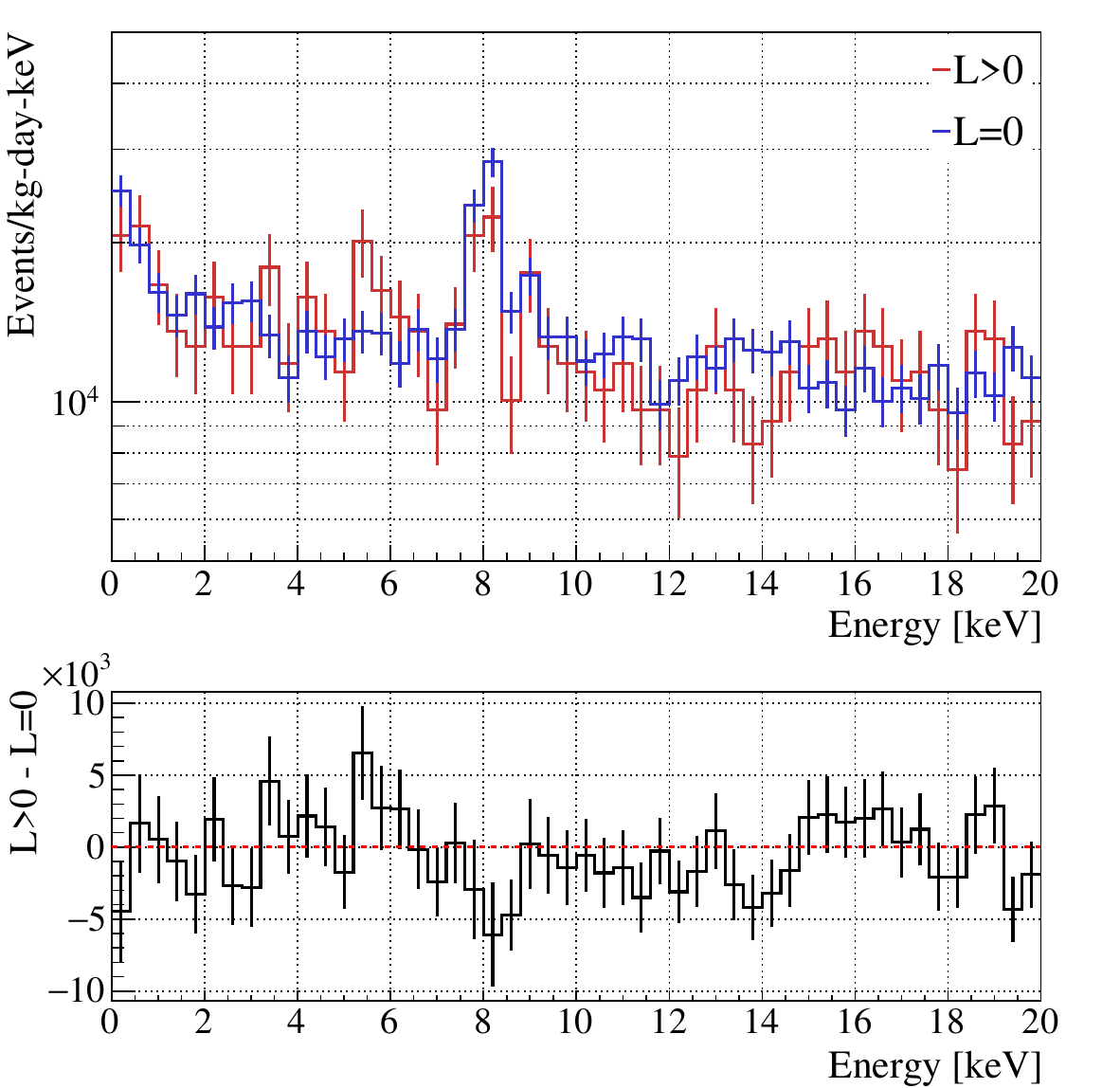}\hfill
\includegraphics[width=0.49\linewidth]{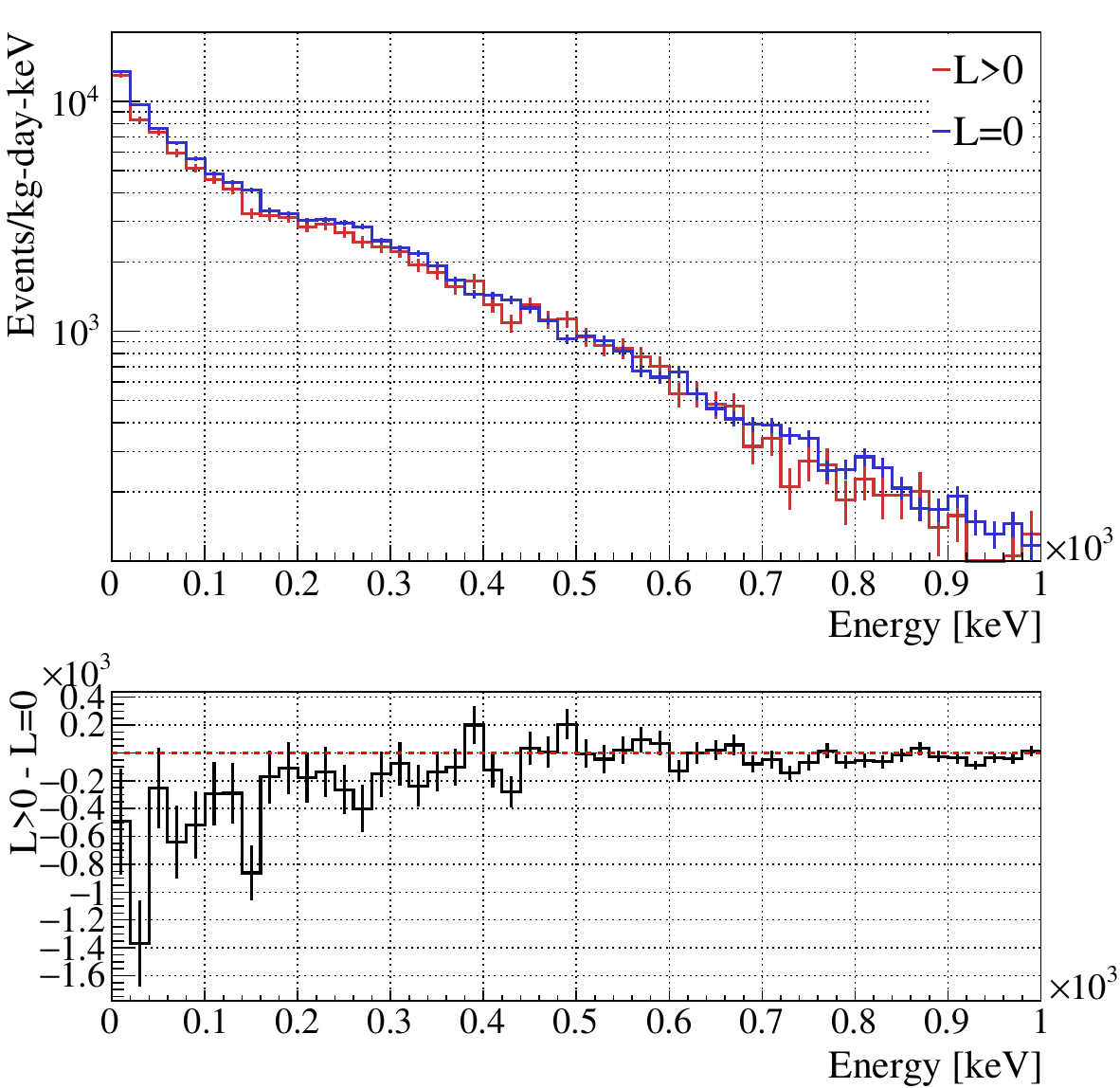}
\caption{Top: High-energy event rates for the Pb-Pb collision period, for images with L=0 (blue) and L$>$0 (red), shown from 0 to 20~keV (left) and from 0 to 1000~keV (right). Bottom: Difference in event rates between images with and without luminosity (black), i.e. L$>$0 - L=0, for the same energy ranges. A higher event rate below $\sim$450~keV for L=0 is observed.}
\label{fig:highEspechi}
\end{figure}

The high-energy event rates for the proton-proton and Pb-Pb collision periods are shown in the top panels of Figures~\ref{fig:highEspecpp} and~\ref{fig:highEspechi}, respectively. We compare the rates corresponding to images with luminosity (L$>$0) and without luminosity (L=0) in two energy reanges, from 0 to 20~keV and from 0 to 1000~keV. In the region below 20~keV, characteristic X-ray emission peaks are visible. The most prominent ones at 8~keV and 8.9~keV correspond to the Cu K$_{\alpha}$ and K$_{\beta}$ emission lines, respectively, from the CCD packaging material. In the spectra up to 1000~keV, an expected bump at $\sim$220~keV corresponding to cosmic muon events is seen.

The differences in event rates between images with and without luminosity, i.e. L$>$0 - L=0, are shown in the bottom panels of Figures~\ref{fig:highEspecpp} and~\ref{fig:highEspechi} for the proton-proton and the Pb-Pb collision periods, respectively. These plots show a higher event rate below $\sim$450~keV for L=0 in both periods, with a more pronounced difference for the Pb-Pb collision period. Above 450~keV, where events are primarily associated to cosmic ray muons, the rates between images with and without luminosity are consistent.

\begin{figure}[ht!]
\centering
\includegraphics[width=0.49\linewidth]{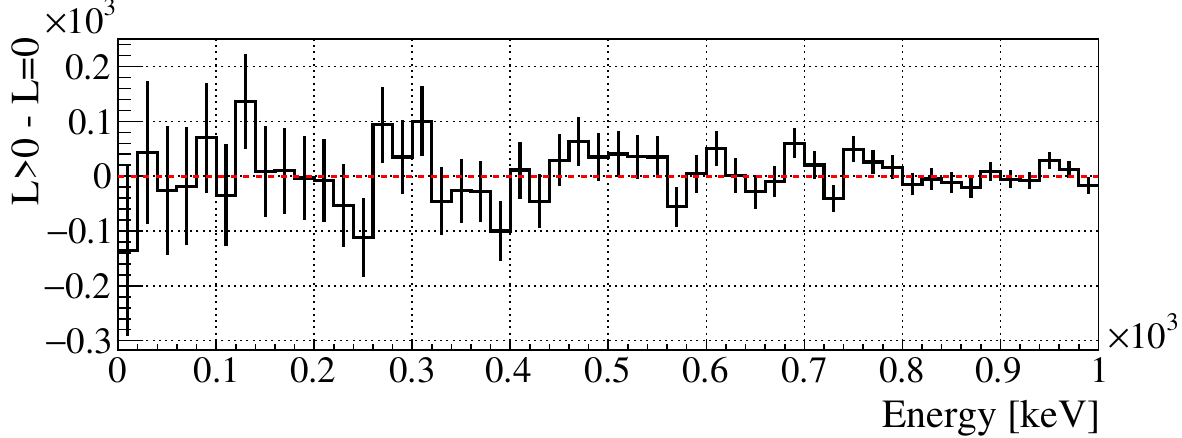}\hfill
\includegraphics[width=0.49\linewidth]{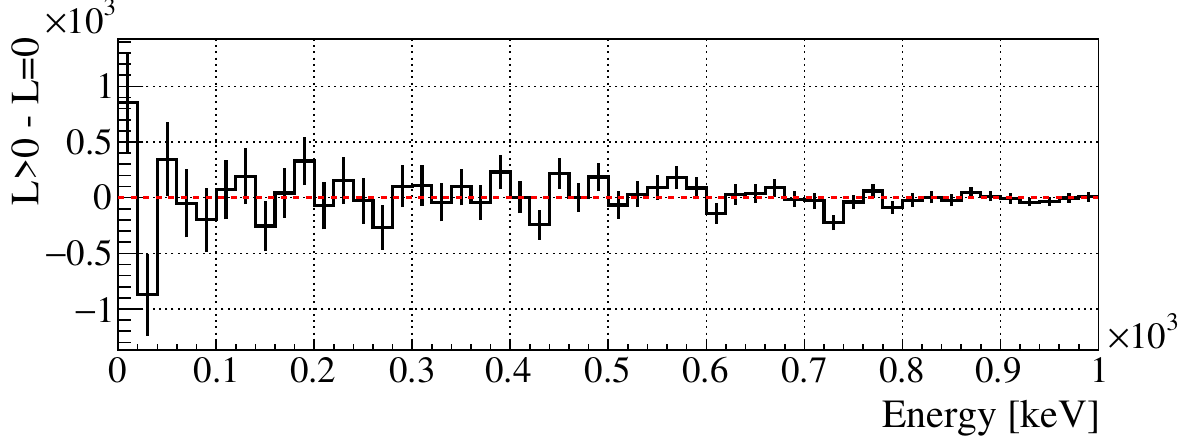}
\caption{Difference in event rates between images with and without luminosity, i.e. L$>$0 - L=0, from 0 to 1000~keV for the proton-proton (left) and the Pb-Pb (right) collision periods, excluding images acquired after the Pb-Pb collisions. Here, the difference is compatible with zero for both periods.}
\label{fig:norun6}
\end{figure}

After a more detailed investigation, we observe an increase in the high-energy event rate beginning after the Pb-Pb collisions and persisting through the end of run 6, a beam-off period. Interestingly, a similar increase is seen in the single-electron rate over the same time frame, see Fig.~\ref{fig:sermoni}. These correlated increases may share a common origin. One possible explanation is the production of secondary neutrons during the Pb-Pb collisions. Such neutrons could generate point defects in the CCD lattice, creating traps whose emission may contribute to the increased SER, while also activating surrounding materials and leading to delayed high-energy events. However, this remains a hypothesis. The origin of the observed higher rate will be investigated in future work. Figure~\ref{fig:norun6} shows the differences in event rates between images with and without luminosity for both collision periods, excluding images acquired after the Pb-Pb collisions. In this case, the inconsistency below $\sim$450~keV vanishes for both periods.

\subsection{Likelihood analysis}~\label{sec:likelihood}
The likelihood function is expressed as the product of Poisson probability distribution functions for the total number of images, $N_{\rm img}$,~\cite{Cowan2011}
\begin{equation}
\mathcal{L}(\mathbf{n}|\mu,\mathbf{b})=\prod_{i=1}^{N_{\rm img}}\frac{e^{-(\mu s_i+b_i)}(\mu s_i+b_i)^{n_i}}{n_i!}\,,
\end{equation}
where $\mathbf{n}=(n_1, n_2, ..., n_{N_{\rm img}})$ with $n_i$ the number of observed events in the $i$-th image, $\mu$ is the signal strength, $\mu s_i$ is the expected number of beam-related events in the $i$-th image, and $\mathbf{b}=(b_1, b_2, ..., b_{N_{\rm img}})$ with $b_i$ the expected number of background events in the $i$-th image. The values of $s_i$ and $b_i$ are given as
\begin{equation}
s_i=\Delta E\times\epsilon_i\times L_i \qquad {\rm and} \qquad b_i=\Delta E\times\epsilon_i \times R_b\,,
\end{equation}
where $\Delta E$ is the energy bin width, $\epsilon_i$ is the effective exposure of the $i$-th image, which includes the bin's detection efficiency from Tables~\ref{tab:lowEpp} and \ref{tab:lowEii}, $L_i$ is the luminosity associated to the $i$-th image and $R_b$ is the background rate computed from images with $L_i=0$.

The profile likelihood ratio is given by
\begin{equation}
\lambda(\mu)=\frac{\mathcal{L}(\mathbf{n}|\mu,\hat{\hat{\mathbf{b}}})}{\mathcal{L}(\mathbf{n}|\hat{\mu},\hat{\mathbf{b}})}\,,
\end{equation}
where $\hat{\hat{\mathbf{b}}}(\mu)$ is the background that maximizes the likelihood function for a given $\mu$, while $\hat{\mu}$ and $\hat{\mathbf{b}}$ are the global maximum estimators of the likelihood function.  

We evaluate the compatibility of the data with different signal strengths using Monte Carlo simulations. For a given $\mu$, we perform many pseudo-experiments. In each, we simulate a full set of images, where the number of events in each image is sampled from a Poisson distribution with mean $\smash{\mu s_i+\hat{\hat{b}}_i(\mu)}$, and compute the test statistic $t_{\mu}=-2\ln\lambda(\mu)$. The many pseudo-experiments yield the $t_{\mu}$ distribution, $f_{\mu}(t)$, from which we compute the p-value as $\smash{p(\mu)=\int_{t_{\rm obs}(\mu)}^{\infty} f_{\mu}(\tilde{t}) d\tilde{t}}$, with $t_{\rm obs}(\mu)$ the test statistic computed from the actual data.

The 95\% C.L. upper limit on the number of events is defined as the expected number of signal events corresponding to the largest value of $\mu$ for which $p(\mu)\geq 0.05$, denoted as $\mu_{95}$. To compute this, we generate many pseudo-experiments. In each, the number of events per image is sampled from a Poisson distribution with mean $\mu_{95}s_i$, and the total number of signal events is obtained by summing over all images. The upper limit is then given by the mean of these total event counts across all pseudo-experiments.

\section{Limits on mCPs}
We compute the expected number of mCPs from proton-proton collisions at 13.6~TeV arriving at MOSKITA, $\phi(\varepsilon,E_{\chi}, m_{\chi})$, using the \hyperlink{https://github.com/GBeauregard/milliqan-pyth-sim}{Pythia8 event generator}~\cite{mCPs_sim_LHC}, previously employed by the milliQan collaboration~\cite{milliQAN2020}. This simulation models mCP production from hadronic decays and their propagation through the 17~m of rock to the drainage gallery, accounting for the CMS magnet's magnetic field. It does not include mCP production via proton bremsstrahlung. For our analysis, we modify the simulation to reflect a luminosity of 113.3 fb$^{-1}$ and the geometry of the MOSKITA sensor.

An mCP reaching the sensor would interact with electrons in the silicon according to the cross section~\cite{Essig2024}
\begin{equation}    
    \label{energy-loss-fermi}
     \frac{d\sigma}{d\omega} = \frac{8\alpha\varepsilon^2}{n_e \beta^2}\int_0^\infty dk \bigg\{\frac{1}{k}\mathrm{Im}\left(-\frac{1}{\tilde{\epsilon}(\omega, k)}\right)+k\left(\beta^2 - \frac{\omega^2}{k^2}\right)\mathrm{Im}\left(\frac{1}{-k^2 + \tilde{\epsilon}(\omega, k)\omega^2}\right)\bigg\}\,,
\end{equation}
where $k=|\mathbf{k}|$ is the mCP three-momentum transfer to the electrons, $\omega$ is the mCP energy transfer to the electrons, $\varepsilon$ is the mCP charge, $\beta=p_{\chi}/E_{\chi}$ is the mCP three-velocity in the detector's rest frame, $n_e$ is the number of electrons per unit volume, $\alpha$ is the fine structure constant and $\tilde{\epsilon}(\omega, k)$ is the dielectric function of the medium. For highly-boosted mCPs, as the ones that could be generated in proton-proton collisions at the LHC, $\beta\simeq1$. To calculate the probability of ionizing a given number of electrons in our sensor, we convolve the cross section in Eq.~\ref{energy-loss-fermi} with the probability of a recoiling electron to produce ionized electrons, as modeled in Ref.~\cite{Ramanathan2020}; this procedure yields the detection cross section, $\sigma_{\rm det}$. The expected number of detected events from mCP interactions is computed as

\begin{equation} \label{eq:evtsfrommcps}
    N(\varepsilon, m_{\chi}) = \epsilon A \int \phi(\varepsilon, E_{\chi}, m_{\chi})\left[1-e^{-dn_e\sigma_{\rm det}}\right]dE_{\chi}\,,
\end{equation}
where $A$ and $d$ are the detector's area and length, respectively, and $\epsilon$ is the effective exposure for the energy range considered~\cite{mCP_by_SENSEI,OIT}.

\begin{figure}[ht!]
    \centering
    \includegraphics[width=0.7\linewidth]{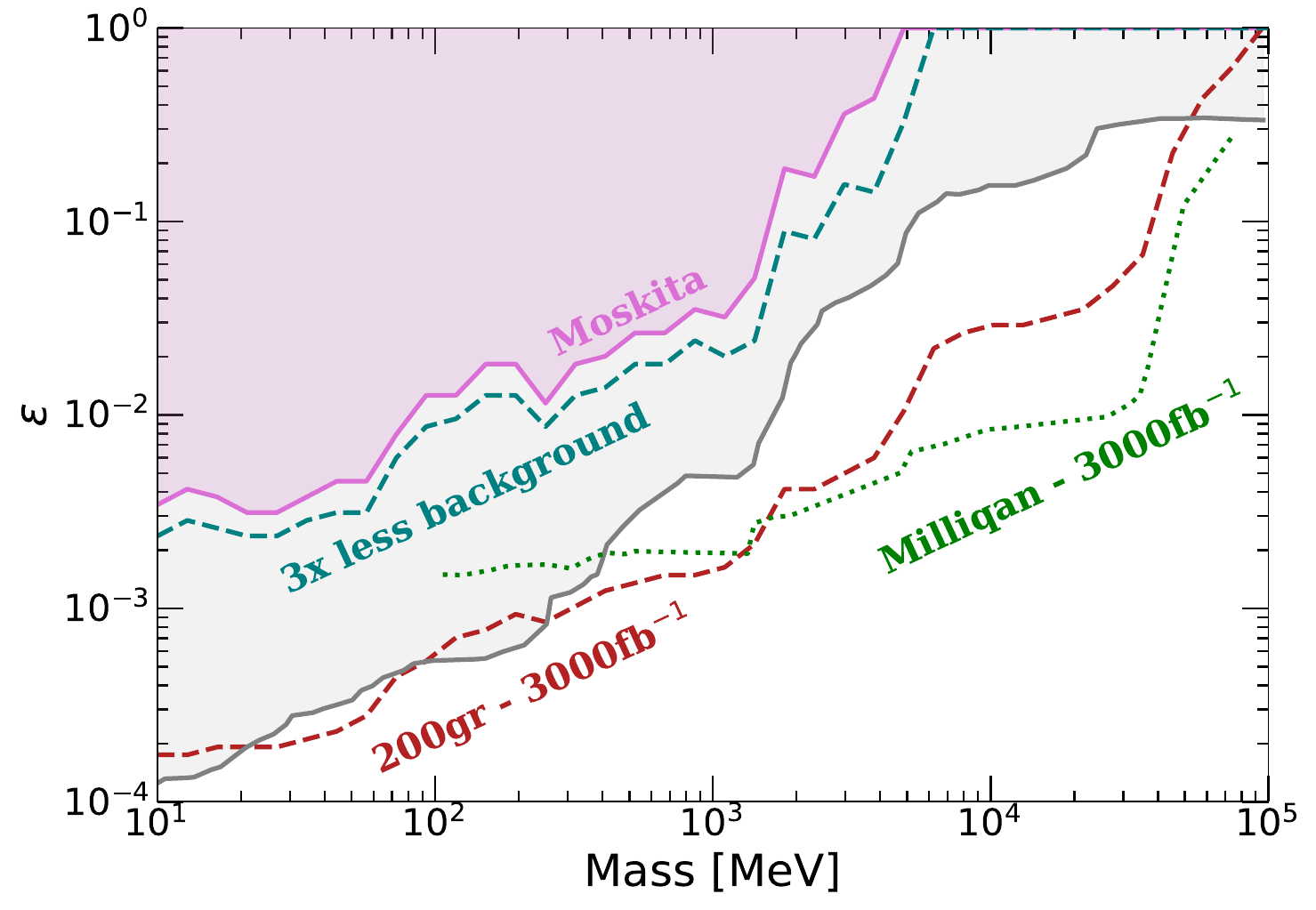}
    \caption{95\% C.L. exclusion region on mCPs from MOSKITA using data collected in 2024 during the LHC proton-proton collisions (pink). Projected 95\% C.L. exclusion regions for: 1) MOSKITA collecting 180~fb$^{-1}$ with 3 times less background (cyan), and 2) a 200~g skipper-CCD detector, with $A=231$~cm$^{2}$ and $d=0.6$~cm, collecting 3000~fb$^{-1}$ with 10 times less background (red). For comparison, constraints from previous mCP searches~\cite{MilliQSLAC, Argoneut_limit, LEP, Magill:2018tbb, milliQAN2020} (gray) are shown, which include SENSEI\cite{mCP_by_SENSEI}, using the same sensor technology, and the most recent results from milliQan~\cite{milliqan2025}, running at the same site. Also, the projected 95\% C.L. exclusion region for the milliQan full bar detector collecting 3000~fb$^{-1}$ during the HL-LHC era~\cite{milliQAN2021} (green) is shown.}
    \label{fig:moskitalim}
\end{figure}

Using Eq.~\ref{eq:evtsfrommcps}, we compute $N(\varepsilon, m_{\chi})$ for the 4$e^-$-7$e^-$ energy range, considering $\epsilon$ as the mean of the effective exposures of the corresponding energy bins in Table~\ref{tab:lowEpp}. The selected energy range corresponds to the region where interactions with bulk plasmons are expected to enhance the signal~\cite{Essig2024} and has been previously used for skipper-CCD mCP searches~\cite{mCP_by_SENSEI,mCP_from_reactors,OIT}. The 95\% C.L. upper limit on the number of signal events in MOSKITA within this energy range, for the proton-proton collision period, computed as described in Section~\ref{sec:likelihood}, is 7.8 events, and the corresponding p-value for testing the null hypothesis, $p(\mu=0)$, is 0.096. The corresponding 95\% C.L. exclusion region in the millicharge-mass parameter space is defined by the condition $N(\varepsilon, m_{\chi})\geq7.8$~events, and is shown in Figure~\ref{fig:moskitalim}. 

We also show in this figure the projected exclusion regions for: 1) MOSKITA collecting 180 fb$^{-1}$ with 3 times less background, and 2) a 200~g skipper-CCD detector, with $A=231$~cm$^{2}$ and $d=0.6$~cm, collecting 3000~fb$^{-1}$ with 10 times less background. Such an increase in mass is plausible as larger, i.e. $\mathcal{O}(100~{\rm g})$, skipper-CCD detectors operating with sub-electron resolution have already been demonstrated~\cite{chierchie2023first}. Similarly, such a background reduction could be achievable by deploying a multi-layered passive shield targeted for external low-energy backgrounds, particularly low-energy neutrons, as those used in skipper-CCD dark matter and reactor experiments~\cite{SENSEIsnolab2025, DAMIC2024, DAMICM2023, CONNIE2024, Atucha2024}. 

\section{Conclusions}
We successfully installed and operated MOSKITA $\sim$33~m from the CMS collision point, the first skipper-CCD detector at the LHC sensitive to low-energy ionization signals from high-energy collisions. Data were acquired throughout 2024, during Run 3 of the LHC, collecting a total luminosity of 113.3~fb$^{-1}$ during the period of proton-proton collisions and 1.54~nb$^{-1}$ during the period of Pb-Pb collisions. Details on the detector installation, performance, data acquisition, and analysis pipeline are provided.

The event rate observed at MOSKITA is reported in two energy regions, $\leq20e^-$ and $>20e^-$, for both collision periods. In the low-energy region ($\leq20e^-$), we perform a likelihood analysis to test for correlations between the event rate and the LHC luminosity. No correlation was found during the period of proton-proton collisions. In contrast, during the Pb-Pb collision period, we observe a significant excess of 7$e^-$ events correlated with beam activity (p=0.017), which may still be consistent with a statistical fluctuation.

Using the computed 95\% C.L. upper limit on the number of 4$e^-$-7$e^-$ signal events in MOSKITA for the proton-proton collision period, and comparing it to the expected number of detected events from mCP interactions, we constrain the millicharge-mass parameter space. While these constraints are not competitive due to the low mass ($\sim$2.2 g) of the MOSKITA sensor, we project that the sensitivity reach of a more massive ($\sim$200~g) skipper-CCD detector for the HL-LHC era would be complementary to those from similar efforts~\cite{milliQAN2021, MoEDAL2024, FASER2025, FORMOSA2025, Flare2023}. It is worth noting that even though these constraints were derived under the mCP model, MOSKITA results could be used to probe other new physics scenarios.

For the high-energy region ($>20e^-$), we find that the measured event rates for images with and without luminosity are inconsistent below 450~keV, for both collision periods. In particular, we observe a higher event rate after the Pb-Pb collisions occurred, coinciding with a rise in the single-electron rate. When excluding images taken after the Pb-Pb collisions, the high-energy event rates for images with and without luminosity become consistent. This increase may be attributed to secondary neutrons produced during the Pb-Pb collisions. However, its origin remains to be investigated in future work.

Overall, the results presented here demonstrate the potential of the skipper-CCD technology to explore new physics at colliders. This work establishes a path forward for future skipper-CCD efforts at high-energy facilities. In particular at the LHC, extending data collection and/or developing a larger skipper-CCD detector will be key to improving sensitivity to rare, low-energy signals.

\section{Acknowledgments}
Authors would like to thank and congratulate the CERN accelerator departments, including engineering and administrative personnel, for the excellent performance of the LHC during Run 3 and for their assistance with the MOSKITA detector installation. We gratefully acknowledge people from the CMS and milliQan collaborations for their support, for sharing space in the drainage gallery, and for the useful discussions held prior to the installation. Finally, authors would also like to thank Gustavo Otero y Garzon and Ricardo Piegaia for the initial talks regarding the possibility of operating a skipper-CCD detector at the LHC complex. This work was done using the resources of the Fermi National Accelerator Laboratory (Fermilab), a U.S. Department of Energy, Office of Science, Office of High Energy Physics HEP User Facility. Fermilab is managed by FermiForward Discovery Group, LLC, acting under Contract No. 89243024CSC000002.

\bibliographystyle{ieeetr}
\bibliography{biblio.bib}

\begin{thebibliography}{10}

\bibitem{Kopp2013}
J.~Kopp, P.~A.~N. Machado, M.~Maltoni, and T.~Schwetz, ``{Sterile neutrino oscillations: the global picture},'' {\em JHEP}, vol.~2013, no.~5, p.~50, 2013.

\bibitem{Schramm:2022ngt}
S.~Schramm, ``{Searching for New Physics in Hadronic Final States with Run 2 Proton\textendash{}Proton Collision Data at the LHC},'' {\em Symmetry}, vol.~14, no.~6, p.~1173, 2022.

\bibitem{Biswas:2022tcw}
S.~Biswas, E.~Gabrielli, and B.~Mele, ``{Dark Photon Searches via Higgs Boson Production at the LHC and Beyond},'' {\em Symmetry}, vol.~14, no.~8, p.~1522, 2022.

\bibitem{MCPdarkPhoton}
B.~Holdom, ``{Two U(1)'s and $\epsilon$ charge shifts},'' {\em Physics Letters B}, vol.~166, no.~2, pp.~196--198, 1986.

\bibitem{Tiffenberg_2017}
J.~Tiffenberg, M.~Sofo-Haro, A.~Drlica-Wagner, R.~Essig, Y.~Guardincerri, S.~Holland, T.~Volansky, and T.-T. Yu, ``{Single-Electron and Single-Photon Sensitivity with a Silicon Skipper CCD},'' {\em Phys. Rev. Lett.}, vol.~119, p.~131802, Sept. 2017.

\bibitem{SENSEI2020}
L.~Barak, I.~M. Bloch, M.~Cababie, G.~Cancelo, L.~Chaplinsky, F.~Chierchie, M.~Crisler, A.~Drlica-Wagner, R.~Essig, J.~Estrada, E.~Etzion, G.~F. Moroni, D.~Gift, S.~Munagavalasa, A.~Orly, D.~Rodrigues, A.~Singal, M.~S. Haro, L.~Stefanazzi, J.~Tiffenberg, S.~Uemura, T.~Volansky, and T.-T. Yu, ``{SENSEI: Direct-Detection Results on sub-GeV Dark Matter from a New Skipper CCD},'' {\em Phys. Rev. Lett.}, vol.~125, p.~171802, Oct. 2020.

\bibitem{mCP_by_SENSEI}
L.~Barak, I.~M. Bloch, A.~M. Botti, M.~Cababie, G.~Cancelo, B.~A. Cervantes-Vergara, L.~Chaplinsky, M.~Crisler, A.~Drlica-Wagner, R.~Essig, J.~Estrada, E.~Etzion, G.~F. Moroni, S.~E. Holland, Y.~Korn, I.~Lawson, S.~Luoma, S.~Munagavalasa, A.~Orly, S.~E. Perez, D.~Rodrigues, N.~A. Saffold, S.~Scorza, A.~Singal, M.~S. Haro, L.~Stefanazzi, K.~Stifter, J.~Tiffenberg, S.~Uemura, T.~Volansky, T.-T. Yu, R.~Harnik, Z.~Liu, and R.~Plestid, ``{Search by the SENSEI Experiment for Millicharged Particles Produced in the NuMI Beam},'' {\em Phys. Rev. Lett.}, vol.~133, p.~071801, Aug. 2024.

\bibitem{mCP_from_reactors}
A.~A. Aguilar-Arevalo, J.~C. D'Olivo, Y.~Sarkis, N.~Avalos, X.~Bertou, P.~Bellino, C.~Bonifazi, A.~Botti, G.~Cancelo, J.~Estrada, R.~Ford, K.~Kuk, A.~Lathrop, J.~Tiffenberg, M.~Cababi\'e, B.~A. Cervantes-Vergara, C.~Chavez, F.~Chierchie, D.~Delgado, E.~Depaoli, J.~a. dos Anjos, H.~P. Lima, G.~F. Moroni, A.~R.~F. Neto, B.~Kilminster, P.~Lemos, K.~Maslova, I.~Nasteva, A.~C. Oliveira, P.~Ventura, M.~Makler, A.~Magnoni, F.~Marinho, J.~Molina, D.~Stalder, S.~Perez, D.~Rodrigues, L.~Paulucci, I.~Sidelnik, and M.~S. Haro, ``{Search for Reactor-Produced Millicharged Particles with Skipper-CCDs at the CONNIE and Atucha-II Experiments},'' {\em Phys. Rev. Lett.}, vol.~134, p.~071801, Feb. 2025.

\bibitem{milliqan2025}
S.~{Alcott}, Z.~{Bhatti}, J.~{Brooke}, C.~{Campagnari}, M.~{Carrigan}, M.~{Citron}, R.~{De Los Santos}, A.~{De Roeck}, C.~{Dorofeev}, T.~{Du}, J.~{Goldstein}, F.~{Golf}, N.~{Gonzalez}, A.~{Haas}, J.~{Heymann}, C.~S. {Hill}, D.~{Imani}, M.~{Joyce}, K.~{Larina}, R.~{Loos}, S.~{Lowette}, H.~{Mei}, D.~W. {Miller}, B.~{Peng}, S.~N. {Santpu}, I.~{Reed}, E.~{Schaffer}, R.~{Schmitz}, J.~{Steenis}, D.~{Stuart}, J.~S. {Tafoya Vargas}, D.~{Vannerom}, T.~{Wybouw}, Z.~{Xiao}, H.~{Zaraket}, G.~{Zecchinelli}, and C.~{Zheng}, ``{Search for millicharged particles in proton-proton collisions at $\sqrt{s} = 13.6$ TeV},'' {\em arXiv e-prints}, June 2025.

\bibitem{MoEDAL2024}
M.~Kalliokoski, V.~A. Mitsou, M.~de~Montigny, A.~Mukhopadhyay, P.-P.~A. Ouimet, J.~Pinfold, A.~Shaa, and M.~Staelens, ``{Searching for minicharged particles at the energy frontier with the MoEDAL-MAPP experiment at the LHC},'' {\em JHEP}, vol.~2024, p.~137, Apr. 2024.

\bibitem{FASER2025}
R.~{Mammen Abraham}, X.~{Ai}, S.~{Alonso-Monsalve}, J.~{Anders}, C.~{Antel}, A.~{Ariga}, T.~{Ariga}, J.~{Atkinson}, F.~U. {Bernlochner}, T.~{Boeckh}, J.~{Boyd}, {\em et~al.}, ``{Prospects and Opportunities with an upgraded FASER Neutrino Detector during the HL-LHC era: Input to the EPPSU},'' {\em arXiv e-prints}, p.~arXiv:2503.19775, Mar. 2025.

\bibitem{FORMOSA2025}
M.~{Citron}, F.~{Golf}, K.~{Gunthoti}, A.~{Haas}, C.~S. {Hill}, D.~{Imani}, S.~{Kelly}, M.~{Liu}, S.~{Lowette}, A.~{De Roeck}, S.~{Neha Santpur}, R.~{Schmitz}, J.~{Steenis}, D.~{Stuart}, Y.-D. {Tsai}, J.~S. {Tafoya Vargas}, T.~{Wybouw}, and J.~{Yoo}, ``{Input to the ESPPU 2026 update: Searching for millicharged particles with the FORMOSA experiment at the CERN LHC},'' {\em arXiv e-prints}, p.~arXiv:2504.12973, Apr. 2025.

\bibitem{Flare2023}
F.~Kling, J.-L. Kuo, S.~Trojanowski, and Y.-D. Tsai, ``{FLArE up dark sectors with EM form factors at the LHC forward physics facility},'' {\em Nuclear Physics B}, vol.~987, p.~116103, 2023.

\bibitem{OscuraSensors2023}
B.~A. Cervantes-Vergara, S.~Perez, J.~Estrada, A.~Botti, C.~R. Chavez, F.~Chierchie, N.~Saffold, A.~Aguilar-Arevalo, F.~Alcalde-Bessia, N.~Avalos, O.~Baez, D.~Baxter, X.~Bertou, C.~Bonifazi, G.~Cancelo, N.~Castelló-Mor, A.~E. Chavarria, J.~M.~D. Egea, J.~C. D'Olivo, C.~Dreyer, A.~Drlica-Wagner, R.~Essig, E.~Estrada, E.~Etzion, P.~Grylls, G.~Fernandez-Moroni, M.~Fernández-Serra, S.~Ferreyra, S.~Holland, A.~L. Barreda, A.~Lathrop, I.~Lawson, B.~Loer, S.~Luoma, E.~M. Villalpando, M.~M. Montero, K.~McGuire, J.~Molina, S.~Munagavalasa, D.~Norcini, A.~Piers, P.~Privitera, D.~Rodrigues, R.~Saldanha, A.~Singal, R.~Smida, M.~Sofo-Haro, D.~Stalder, L.~Stefanazzi, J.~Tiffenberg, M.~Traina, S.~Uemura, P.~Ventura, R.~V. Cortabitarte, and R.~Yajur, ``{Skipper-CCD sensors for the Oscura experiment: requirements and preliminary tests},'' {\em JINST}, vol.~18, p.~P08016, Aug. 2023.

\bibitem{PM2023}
B.~Cervantes-Vergara, S.~Perez, J.~D’Olivo, J.~Estrada, D.~Grimm, S.~Holland, M.~Sofo-Haro, and W.~Wong, ``{Skipper-CCDs: Current applications and future},'' {\em Nuclear Instruments and Methods in Physics Research Section A: Accelerators, Spectrometers, Detectors and Associated Equipment}, vol.~1046, p.~167681, 2023.

\bibitem{milliQAN2021}
A.~Ball, J.~Brooke, C.~Campagnari, M.~Carrigan, M.~Citron, A.~De~Roeck, M.~Ezeldine, B.~Francis, M.~Gastal, M.~Ghimire, J.~Goldstein, F.~Golf, A.~Haas, R.~Heller, C.~S. Hill, L.~Lavezzo, R.~Loos, S.~Lowette, B.~Manley, B.~Marsh, D.~W. Miller, B.~Odegard, R.~Schmitz, F.~Setti, H.~Shakeshaft, D.~Stuart, M.~Swiatlowski, J.~Yoo, and H.~Zaraket, ``{Sensitivity to millicharged particles in future proton-proton collisions at the LHC with the milliQan detector},'' {\em Phys. Rev. D}, vol.~104, p.~032002, Aug. 2021.

\bibitem{milliQAN2020}
A.~Ball, G.~Beauregard, J.~Brooke, C.~Campagnari, M.~Carrigan, M.~Citron, J.~De~La~Haye, A.~De~Roeck, Y.~Elskens, R.~E. Franco, M.~Ezeldine, B.~Francis, M.~Gastal, M.~Ghimire, J.~Goldstein, F.~Golf, J.~Guiang, A.~Haas, R.~Heller, C.~S. Hill, L.~Lavezzo, R.~Loos, S.~Lowette, G.~Magill, B.~Manley, B.~Marsh, D.~W. Miller, B.~Odegard, F.~R. Saab, J.~Sahili, R.~Schmitz, F.~Setti, H.~Shakeshaft, D.~Stuart, M.~Swiatlowski, J.~Yoo, H.~Zaraket, and H.~Zheng, ``{Search for millicharged particles in proton-proton collisions at $\sqrt{s}=13\text{ }\text{ }\mathrm{TeV}$},'' {\em Phys. Rev. D}, vol.~102, p.~032002, Aug. 2020.

\bibitem{CMS2008}
S.~Chatrchyan, G.~Hmayakyan, V.~Khachatryan, A.~M. Sirunyan, W.~Adam, T.~Bauer, T.~Bergauer, H.~Bergauer, M.~Dragicevic, J.~Erö, {\em et~al.}, ``{The CMS experiment at the CERN LHC},'' {\em JINST}, vol.~3, p.~S08004, Aug. 2008.

\bibitem{SENSEIsnolab2025}
P.~Adari, I.~M. Bloch, A.~M. Botti, M.~Cababie, G.~Cancelo, B.~A. Cervantes-Vergara, M.~Crisler, M.~Daal, A.~Desai, A.~Drlica-Wagner, R.~Essig, J.~Estrada, E.~Etzion, G.~F. Moroni, S.~E. Holland, J.~Kehat, Y.~Korn, I.~Lawson, S.~Luoma, A.~Orly, S.~E. Perez, D.~Rodrigues, N.~A. Saffold, S.~Scorza, A.~Singal, M.~Sofo-Haro, L.~Stefanazzi, K.~Stifter, J.~Tiffenberg, S.~Uemura, E.~M. Villalpando, T.~Volansky, Y.~Wu, T.-T. Yu, T.~Emken, and H.~Xu, ``{First Direct-Detection Results on Sub-GeV Dark Matter Using the SENSEI Detector at SNOLAB},'' {\em Phys. Rev. Lett.}, vol.~134, p.~011804, Jan. 2025.

\bibitem{LTA2021}
G.~I. Cancelo, C.~Chavez, F.~Chierchie, J.~Estrada, G.~Fernandez-Moroni, E.~E. Paolini, M.~S. Haro, A.~Soto, L.~Stefanazzi, J.~Tiffenberg, K.~Treptow, N.~Wilcer, and T.~J. Zmuda, ``{Low threshold acquisition controller for Skipper charge-coupled devices},'' {\em JATIS}, vol.~7, no.~1, p.~015001, 2021.

\bibitem{Cowan2011}
G.~Cowan, K.~Cranmer, E.~Gross, and O.~Vitells, ``{Asymptotic formulae for likelihood-based tests of new physics},'' {\em The European Physical Journal C}, vol.~71, p.~1554, Feb. 2011.

\bibitem{mCPs_sim_LHC}
G.~Beauregard, ``{Pythia mCP Simulation}.'' \url{https://github.com/GBeauregard/milliqan-pyth-sim}, 2018.

\bibitem{Essig2024}
R.~Essig, R.~Plestid, and A.~Singal, ``{Collective excitations and low-energy ionization signatures of relativistic particles in silicon detectors},'' {\em Communications Physics}, vol.~7, p.~416, Dec. 2024.

\bibitem{Ramanathan2020}
K.~Ramanathan and N.~Kurinsky, ``{Ionization yield in silicon for eV-scale electron-recoil processes},'' {\em Phys. Rev. D}, vol.~102, p.~063026, Sept. 2020.

\bibitem{OIT}
S.~Perez, D.~Rodrigues, J.~Estrada, R.~Harnik, Z.~Liu, B.~A. Cervantes-Vergara, J.~C. D'Olivo, R.~D. Plestid, J.~Tiffenberg, T.-T. Yu, A.~Aguilar-Arevalo, F.~Alcalde-Bessia, N.~Avalos, O.~Baez, D.~Baxter, X.~Bertou, C.~Bonifazi, A.~Botti, G.~Cancelo, N.~Castell{\'o}-Mor, A.~E. Chavarria, C.~R. Chavez, F.~Chierchie, J.~M. De~Egea, C.~Dreyer, A.~Drlica-Wagner, R.~Essig, E.~Estrada, E.~Etzion, P.~Grylls, G.~Fernandez-Moroni, M.~Fern{\'a}ndez-Serra, S.~Ferreyra, S.~Holland, A.~L. Barreda, A.~Lathrop, I.~Lawson, B.~Loer, S.~Luoma, E.~M. Villalpando, M.~M. Montero, K.~McGuire, J.~Molina, S.~Munagavalasa, D.~Norcini, A.~Piers, P.~Privitera, N.~Saffold, R.~Saldanha, A.~Singal, R.~Smida, M.~Sofo-Haro, D.~Stalder, L.~Stefanazzi, M.~Traina, Y.-D. Tsai, S.~Uemura, P.~Ventura, R.~V. Cortabitarte, and R.~Yajur, ``{Searching for millicharged particles with 1 kg of Skipper-CCDs using the NuMI beam at Fermilab},'' {\em JHEP}, vol.~2024, p.~72, Feb. 2024.

\bibitem{chierchie2023first}
F.~Chierchie, C.~Chavez, M.~Sofo~Haro, G.~Fernandez~Moroni, B.~Cervantes-Vergara, S.~Perez, J.~Estrada, J.~Tiffenberg, S.~Uemura, and A.~Botti, ``{First results from a multiplexed and massive instrument with sub-electron noise Skipper-CCDs},'' {\em JINST}, vol.~18, p.~P01040, Jan. 2023.

\bibitem{DAMIC2024}
A.~Aguilar-Arevalo, I.~Arnquist, N.~Avalos, L.~Barak, D.~Baxter, X.~Bertou, I.~M. Bloch, A.~M. Botti, M.~Cababie, G.~Cancelo, N.~Castell\'o-Mor, B.~A. Cervantes-Vergara, A.~E. Chavarria, J.~Cortabitarte-Guti\'errez, M.~Crisler, J.~Cuevas-Zepeda, A.~Dastgheibi-Fard, C.~De~Dominicis, O.~Deligny, A.~Drlica-Wagner, J.~Duarte-Campderros, J.~C. D'Olivo, R.~Essig, E.~Estrada, J.~Estrada, E.~Etzion, F.~Favela-Perez, N.~Gadola, R.~Ga\"{\i}or, S.~E. Holland, T.~Hossbach, L.~Iddir, B.~Kilminster, Y.~Korn, A.~Lantero-Barreda, I.~Lawson, S.~Lee, A.~Letessier-Selvon, P.~Loaiza, A.~Lopez-Virto, S.~Luoma, E.~Marrufo-Villalpando, K.~J. McGuire, G.~F. Moroni, S.~Munagavalasa, D.~Norcini, A.~Orly, G.~Papadopoulos, S.~Paul, S.~E. Perez, A.~Piers, P.~Privitera, P.~Robmann, D.~Rodrigues, N.~A. Saffold, S.~Scorza, M.~Settimo, A.~Singal, R.~Smida, M.~Sofo-Haro, L.~Stefanazzi, K.~Stifter, J.~Tiffenberg, M.~Traina, S.~Uemura, I.~Vila, R.~Vilar, T.~Volansky, G.~Warot, R.~Yajur, T.-T. Yu, and J.-P. Zopounidis, ``{Confirmation of the
  spectral excess in DAMIC at SNOLAB with skipper CCDs},'' {\em Phys. Rev. D}, vol.~109, p.~062007, Mar. 2024.

\bibitem{DAMICM2023}
I.~Arnquist, N.~Avalos, D.~Baxter, X.~Bertou, N.~Castell\'o-Mor, A.~E. Chavarria, J.~Cuevas-Zepeda, J.~C. Guti\'errez, J.~Duarte-Campderros, A.~Dastgheibi-Fard, O.~Deligny, C.~De~Dominicis, E.~Estrada, N.~Gadola, R.~Ga\"{\i}or, T.~Hossbach, L.~Iddir, L.~Khalil, B.~Kilminster, A.~Lantero-Barreda, I.~Lawson, S.~Lee, A.~Letessier-Selvon, P.~Loaiza, A.~Lopez-Virto, A.~Matalon, S.~Munagavalasa, K.~J. McGuire, P.~Mitra, D.~Norcini, G.~Papadopoulos, S.~Paul, A.~Piers, P.~Privitera, K.~Ramanathan, P.~Robmann, M.~Settimo, R.~Smida, R.~Thomas, M.~Traina, I.~Vila, R.~Vilar, G.~Warot, R.~Yajur, and J.-P. Zopounidis, ``{First Constraints from DAMIC-M on Sub-GeV Dark-Matter Particles Interacting with Electrons},'' {\em Phys. Rev. Lett.}, vol.~130, p.~171003, Apr. 2023.

\bibitem{CONNIE2024}
A.~A. {Aguilar-Arevalo}, N.~{Avalos}, X.~{Bertou}, C.~{Bonifazi}, G.~{Cancelo}, B.~A. {Cervantes-Vergara}, C.~{Chavez}, F.~{Chierchie}, G.~{Coelho Corr{\^e}a}, J.~C. {D'Olivo}, J.~{dos Anjos}, J.~{Estrada}, G.~{Fernandez Moroni}, A.~R. {Fernandes Neto}, R.~{Ford}, B.~{Kilminster}, K.~{Kuk}, A.~{Lathrop}, P.~{Lemos}, H.~P. {Lima}, Jr., M.~{Makler}, K.~{Maslova}, F.~{Marinho}, J.~{Molina}, I.~{Nasteva}, A.~C. {Oliveira}, L.~{Paulucci}, D.~{Rodrigues}, Y.~{Sarkis}, M.~{Sofo-Haro}, D.~{Stalder}, J.~{Tiffenberg}, S.~{Uemura}, and P.~{Ventura}, ``{Searches for CE$\nu$NS and Physics beyond the Standard Model using Skipper-CCDs at CONNIE},'' {\em arXiv e-prints}, p.~arXiv:2403.15976, Mar. 2024.

\bibitem{Atucha2024}
E.~Depaoli, D.~Rodrigues, I.~Sidelnik, P.~Bellino, A.~Botti, D.~Delgado, M.~Cababi{\'e}, F.~Chierchie, J.~Estrada, G.~F. Moroni, S.~Perez, and J.~Tiffenberg, ``{Deployment and performance of a Low-Energy-Threshold Skipper-CCD inside a nuclear reactor},'' {\em JHEP}, vol.~2024, p.~155, Oct. 2024.

\bibitem{MilliQSLAC}
A.~A. Prinz, R.~Baggs, J.~Ballam, S.~Ecklund, C.~Fertig, J.~A. Jaros, K.~Kase, A.~Kulikov, W.~G.~J. Langeveld, R.~Leonard, T.~Marvin, T.~Nakashima, W.~R. Nelson, A.~Odian, M.~Pertsova, G.~Putallaz, and A.~Weinstein, ``{Search for Millicharged Particles at SLAC},'' {\em Phys. Rev. Lett.}, vol.~81, pp.~1175--1178, Aug. 1998.

\bibitem{Argoneut_limit}
R.~Acciarri, C.~Adams, J.~Asaadi, B.~Baller, T.~Bolton, C.~Bromberg, F.~Cavanna, D.~Edmunds, R.~S. Fitzpatrick, B.~Fleming, R.~Harnik, C.~James, I.~Lepetic, B.~R. Littlejohn, Z.~Liu, X.~Luo, O.~Palamara, G.~Scanavini, M.~Soderberg, J.~Spitz, A.~M. Szelc, W.~Wu, and T.~Yang, ``{Improved Limits on Millicharged Particles Using the ArgoNeuT Experiment at Fermilab},'' {\em Phys. Rev. Lett.}, vol.~124, p.~131801, Apr. 2020.

\bibitem{LEP}
S.~Davidson, S.~Hannestad, and G.~Raffelt, ``{Updated bounds on milli-charged particles},'' {\em JHEP}, vol.~2000, p.~003, May 2000.

\bibitem{Magill:2018tbb}
G.~Magill, R.~Plestid, M.~Pospelov, and Y.-D. Tsai, ``{Millicharged Particles in Neutrino Experiments},'' {\em Phys. Rev. Lett.}, vol.~122, p.~071801, Feb. 2019.

\end{thebibliography}

\end{document}